\documentclass{article}

\usepackage{arxiv}

\usepackage[utf8]{inputenc} 
\usepackage[T1]{fontenc}    
\usepackage{hyperref}       
\usepackage{url}            
\usepackage{booktabs}       
\usepackage{amsfonts}       
\usepackage{eqnarray,amsmath} 
\usepackage{nicefrac}       
\usepackage{microtype}      
\usepackage{lipsum}		
\usepackage{graphicx}
\usepackage[caption = false]{subfig}
\usepackage{float}
\usepackage{xurl}
\usepackage{doi}
\usepackage{natbib}
\setcitestyle{authoryear,open={(},close={)}} 

\usepackage{xr}
\makeatletter
\newcommand*{\addFileDependency}[1]{
\typeout{(#1)}
\@addtofilelist{#1}
\IfFileExists{#1}{}{\typeout{No file #1.}}
}
\makeatother
\usepackage{color}

\newcommand{\degree}[1]{^{\circ}\,\mathrm{#1}}

\title{Intraseasonal synchronization of extreme rainfalls between North India and the Sahel}
\renewcommand{\shorttitle}{}

\date{\today} 					
\author{
    Felix M. Strnad \\
    Machine Learning in Climate Science, University of Tübingen, Germany \\
    \AND
    Kieran M. R. Hunt\\
    Department of Meteorology,
    University of Reading, UK\\
    National Centre for Atmospheric Science,
    University of Reading, UK\\
    \AND
    Niklas Boers\\
    Earth System Modelling, School of Engineering \& Design, Technical University of Munich, Germany \\
    Potsdam Institute for Climate Impact Research, Potsdam, Germany\\
    Department of Mathematics and Global Systems Institute, University of Exeter, Exeter, UK\\
    \AND
    Bedartha Goswami\\
    Machine Learning in Climate Science,
    University of Tübingen, Germany \\
}

\hypersetup{
pdftitle={Synchronization},
pdfauthor={Felix Strnad},
pdfkeywords={},
}
\usepackage{verbatim}

\begin{document}

\maketitle
\begin{abstract}
    The Indian Summer Monsoon (ISM) and the West African Monsoon (WAM) are dominant drivers of boreal summer precipitation variability in tropical and subtropical regions.
    Although the regional precipitation dynamics in these two regions have been extensively studied, the intraseasonal interactions between the ISM and WAM remain poorly understood.
    Here, we employ a climate network approach based on extreme rainfall events to uncover synchronously occurring extreme rainfall patterns across the two monsoon systems.
    We reveal strong synchronization of extreme rainfall events during the peak monsoon period in July and August, linking heavy rainfall over North India to that over the Sahel with a lag of around 12 days.
    We find that La Ni\~na-like conditions in combination with the Boreal Summer Intraseasonal Oscillation and an enhanced Tropical Easterly Jet (TEJ) foster the synchronization between the ISM and the WAM.
    Convective clouds are transported by an intensified TEJ from southwestern Asia toward North Africa, supporting anomalous deep convection over the Sahel region.
\end{abstract}
Key points:
\begin{itemize}
    \item Climate networks reveal synchronous occurrence of extreme rainfall events in North India and the Sahel.
    \item Propagation pathways leading to the synchronization are identified using multivariate latent space clustering.
    \item The synchronization is driven by a combination of active BSISO, La Ni\~na-like conditions, and a strong Tropical Easterly Jet.
\end{itemize}
\keywords{Communities of rainfall extremes \and Long-range rainfall teleconnection \and intraseasonal rainfall dynamics}

\section{Introduction}
The Indian Summer Monsoon (ISM) and the West African Monsoon (WAM) are primary drivers of boreal summer precipitation variability in tropical and subtropical regions.
The ISM is characterized by a strong meridional overturning circulation, with a low-level monsoon flow from the Indian Ocean to the Indian subcontinent and a return flow at upper levels \citep{Bordoni2008}.
Similarly, the WAM is characterized by a low-level inflow of moist air from the Atlantic Ocean, which converges with the dry, hot air from the Sahara desert.
Both the North Indian and Sahel regions exhibit substantial variations in monsoon rainfall known as ``active'' and ``break'' phases. The active phase is often associated with extreme rainfall events, which can have severe socioeconomic impacts \citep{Kotz2022}.

The intraseasonal variability comes about through a complex interplay of tropical and subtropical large-scale modes of variability \citep{DiCapua2020a}.
For example, the rainfall dynamics in both North India and the Sahel are known to be correlated to the intensity and location of the Tropical Easterly Jet (TEJ) \citep{Huang2019}.
The TEJ is a strong zonal wind maximum in the upper troposphere that is located over the tropical Indian Ocean and is mainly present during boreal summer months (usually June--September).
It is established due to the meridional temperature difference between the Equatorial Indian Ocean and the Tibetan Plateau and extends from the Pacific Ocean to the West Coast of the Sahel Zone (SZ) \citep{Koteswaram1958}.
Previous studies have found a strong connection between the strength of the TEJ and ISM rainfall \citep{Pattanaik2000, Madhu2014, Huang2019, Huang2021}. Through latent heat release, the TEJ becomes stronger (weaker) during wet (dry) years of the ISM \citep{Sathiyamoorthy2007, Rao2016}. Conversely, a stronger upper-level jet increases the vertical wind shear leading to more convective organization and consequently to more rainfall over India.
The physical mechanism underlying the link between the TEJ and Sahel rainfall is less well understood, even though the strong statistical correlation between the TEJ and Sahel rainfall is well documented \citep{Lemburg2019, Nicholson2021}.
For instance, several studies have shown that wet years in the Sahel are often characterized by a regionally stronger TEJ even on interannual timescales \citep{Lemburg2019}.
A strong (weak) TEJ is conducive to wet (dry) conditions and recent work has also shown that there is only a weak relationship between TEJ variability and daily variations in convection over the Sahel \citep{Nicholson2021}. The authors conclude that the intensity of the TEJ is not influenced by rainfall variations over North West and Central Africa.

While the TEJ is influenced by strong convection over India, the occurrence of extreme rainfall events in the Indo-Pacific region itself is substantially influenced by another large-scale mode of variability: the Boreal Summer Intraseasonal Oscillation (BSISO) \citep{Kikuchi2021}.
The BSISO is a mode of large-scale convective variability that originates in the Indian Ocean and propagates northeastward across the Maritime Continent and the Western Pacific \citep{Kikuchi2012, Lee2013, Kiladis2014}.
This oscillation plays a crucial role in the generation of deep convection and interacts with the local circulation patterns over the Indian subcontinent, e.g. by modulating the frequency and behavior of low-pressure systems \citep{Hunt2022}.
The ISM rainfall dynamic is further modulated, on interannual timescales, by the El Ni\~no Southern Oscillation (ENSO).
Indian rainfall is enhanced (reduced) during La Ni\~na (El Ni\~no) years, typically explained by the modulation of the Walker circulation \citep{Kumar2006}.
The BSISO itself is modulated by ENSO with La Ni\~na-like conditions leading to an intensified northward propagation and more EREs in North India \citep{Strnad2023}.
Similarly, due to constructive (destructive) inference from the Walker circulation in La Ni\~na (El Ni\~no) years, the TEJ intensifies (weakens), and expands both horizontally and vertically (contracts spatially) \citep{Nithya2017}.
The modulation of the TEJ by ENSO is used to explain why El Ni\~no events reduce rainfall and foster the occurrence of droughts in Ethiopia \citep{Gleixner2017}. This teleconnection mechanism is also present in some global circulation models \citep{Vashisht2021}.

Even though the two monsoon systems of the ISM and WAM are both influenced by the same global large-scale modes of variability, the occurrence of rainfall in the two monsoon systems of the ISM and WAM is commonly still thought of as being interconnected only on interannual to decadal timescales in the framework of the `Global Monsoon' \citep{Wang2008, Geen2020}.
Possible interactions between both monsoon systems on intraseasonal timescales, however, have not been systematically investigated.
In their study, \cite{Boers2019} reveal a synchronization pattern of extreme rainfall events (EREs) between North India and the Sahel.
While some of the other teleconnections of the ISM shown in \citet{Boers2019} have been explained in recent years; for example, the dynamical mechanism linking the ISM to the circumglobal teleconnection \citep{Beverley2019, Beverley2021} or to the Yellow River Basin \citep{Gupta2022}, the reasons for the synchronization between the ISM and WAM remain unclear.
Our study thus aims to identify the atmospheric processes that drive the synchronization between the ISM and the WAM.

Previous studies on rainfall variability during boreal summer monsoon have traditionally used methods such as linear regression, empirical orthogonal functions (EOFs), and composite analyses \citep{Ding2005, Ding2007, Wang2008, Vellore2014, Walker2015}. These are not suitable to capture the spatial characteristics of EREs.
Additionally, the identification of specific and possibly varying time lags associated with particular interaction patterns of different atmospheric processes is challenging.
Therefore, in this study, we examine the spatial patterns associated with EREs that occur concurrently in the tropics and subtropics by employing a climate network approach \citep{Tsonis2008, Malik2010, Boers2014, Boers2019, Strnad2022, Strnad2023}.
Climate networks are networks where links reflect a strong statistical correlation between corresponding nodes, typically representing time series from different spatial locations. Here, the correlation is accessed by quantifying the synchronicity of EREs using the event synchronization algorithm \citep{Quiroga2002}.
We adopt the methodology introduced in \cite{Strnad2023} to identify groups of densely connected nodes in the network, referred to as communities. These can be understood as spatial regions where EREs occur significantly synchronously.

Our method uncovers a community of synchronous EREs comprising North India, China, and the Sahel region.
We then investigate the mechanisms responsible for the synchronization by building upon the method from \citet{Schloer2023}, clustering the propagation of EREs in a multivariate latent space.
Our approach reveals a robust time-lagged connection between the ISM and WAM monsoon regions, prompting a subsequent exploration using traditional meteorological analysis tools.
We can show that the aforementioned large-scale modes of variability, namely the BSISO, the TEJ, and ENSO interact to foster the synchronization between the ISM and the WAM.

\section{Data and Methods} \label{sec:methods}
\subsection{Datasets used}\label{sec:data}
\paragraph{Precipitation data}
We use the $0.25^{\circ}$ daily resolved precipitation data from the Multi-Source Weighted-Ensemble Precipitation (MSWEP) dataset \citep{Beck2019} for the period of 1980--2022.
The MSWEP dataset is chosen for its longer time range compared to other available multi-satellite precipitation products, improving statistical robustness. It has been shown to represent high rainfall quantiles well on global scales \citep{Beck2019b}.
We restrict our analysis to the tropics and the subtropics of the Northern Hemisphere ($180\degree{W}$ \--- $180\degree{E}$, $30\degree{S}$ \--- $70\degree{N}$, see Fig.~\ref{fig:cd_sketch}~a).

\paragraph{Reanalysis data}
To analyze large-scale patterns associated with the synchronization we use the following variables from the ERA5 Global Reanalysis dataset \citep{Hersbach2020}: Daily outgoing longwave radiation (OLR), sea surface temperature (SST), and on pressure levels from $50$--$1000$~hPa horizontal $(u,v)$-wind fields, vertical velocity $w$, specific humidity  $q$ and temperature $T$.
The datasets used for the multivariate PCA analysis are interpolated onto a $1^{\circ}\times1^{\circ}$ grid, and all remaining datasets are interpolated onto a $2.5^{\circ} \times 2.5^{\circ}$ grid.
To ensure the robustness of our analysis and to avoid confounding effects from long-term climate change, we apply a linear detrending to all reanalysis datasets.

\paragraph{ENSO index}
The ENSO state is obtained using the Multivariate ENSO index version 2 (MEIv2)  \citep{Wolter2011}.
The MEIv2 is a multivariate index that combines six observed variables over the tropical Pacific Ocean to provide a comprehensive measure of the ENSO state. The MEIv2 is available from the  NOAA Physical Sciences Laboratory (PSL) at \url{https://psl.noaa.gov/enso/mei/} (Last Accessed: 10th April 2024).

\paragraph{BSISO index}
The daily resolved BSISO index by \cite{Kikuchi2012} is taken from \url{https://iprc.soest.hawaii.edu/users/kazuyosh/Bimodal_ISO.html} (Last Accessed: 10th May 2023).
The first two leading principal components (PCs) are used to define the state of the BSISO by the amplitude $A=\sqrt{PC_1^2 + PC_2^2}$, where $A\geq 1$ ($A<1$) is regarded as active (inactive) \citep{Wheeler2004}.
The two-dimensional phase space spanned by $PC_1$ and $PC_2$ is subdivided into eight equally sized sections that denote the phase of the BSISO.

\paragraph{Tropical Easterly Jet index}
We employ the index definition of \cite{Huang2019} to describe the characteristics in terms of amplitude and variability of the TEJ for the period from June through September.
This index, denoted as Tropical Easterly Jet Index ($\text{TEJI}$), is mathematically formalized as the mean anomaly time series of the zonal winds ($u$) at the $200$~hPa pressure level, of locations in the box within the interval [$0\degree{E}, 70\degree{E}$], [$0\degree{N}, 15\degree{N}$] (magenta boxes in SI Fig.~\ref{fig:def_tej_pattern}).
The metric of $\text{TEJI}$ is computed by spatially averaging the $u$-wind anomalies enclosed within these boxes. We define positive and negative phases of the $\text{TEJI}$ as events exceeding one standard deviation from the mean (below for negative, above for positive; see SI Fig.~\ref{fig:def_tej_pattern}).


\paragraph{PDO index}
The Pacific Decadal Oscillation (PDO) is a long-term climate pattern that is characterized by a positive and negative phase (Fig.~\ref{fig:PDO_phases}). The PDO is known to be connected with ENSO and influences the climate in the Indo-Pacific domain and the North American continent \citep{DiLorenzo2023}. The PDO index is defined as the first EOF of the monthly SST anomalies in the North Pacific Ocean. It is obtained from the  NOAA Physical Sciences Laboratory (PSL) at \url{https://psl.noaa.gov/data/climateindices/list/} (Last Accessed: 26th April 2024).

\subsection{Communities of Synchronous Extreme Rainfall Events} \label{sec:community_detection}
\begin{figure}[!htb]
    \centering
    \includegraphics[width=1.0\textwidth]{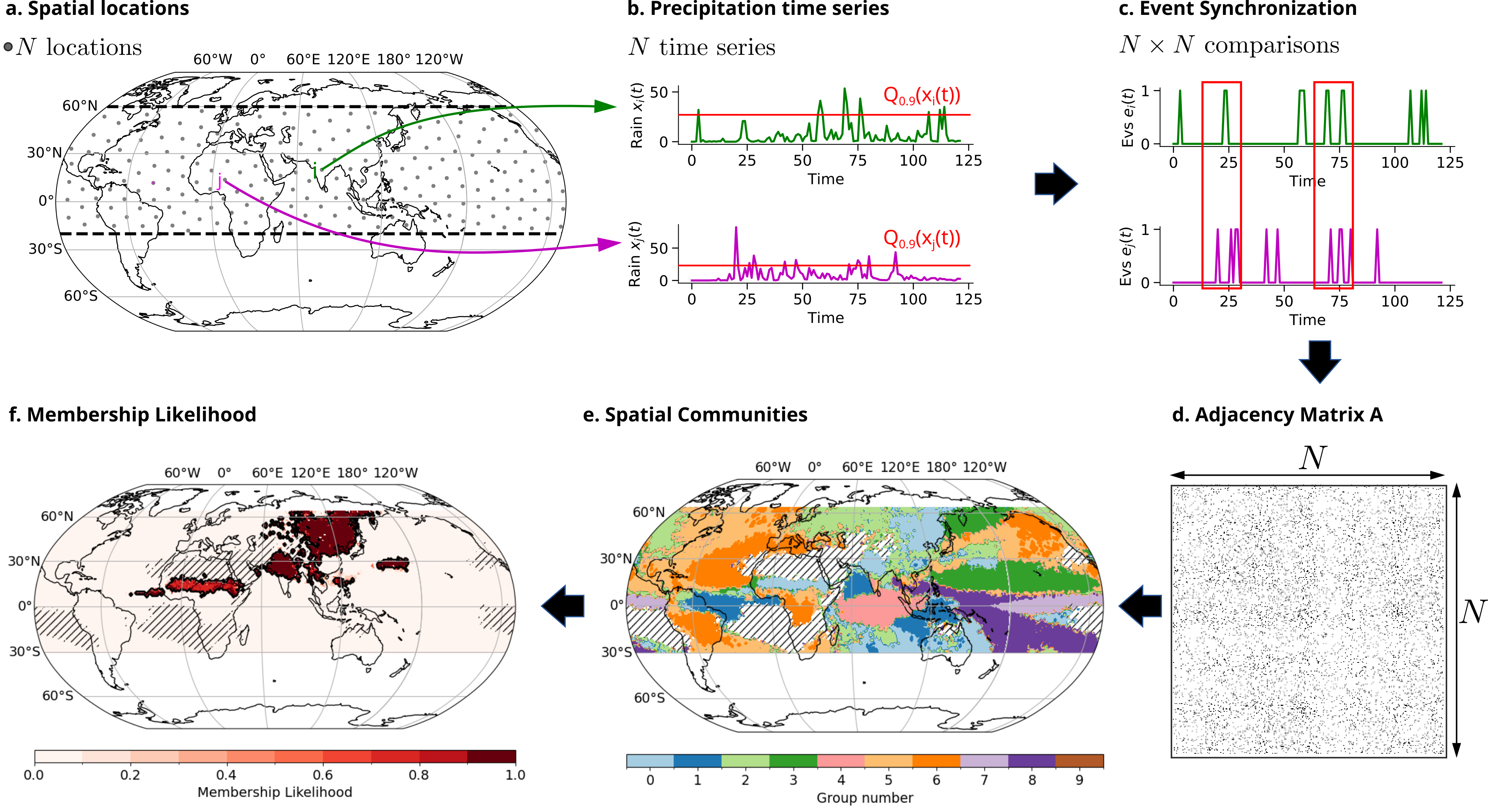}
    \caption{\textbf{Schematic of the Community Detection approach.}
        \textbf{a} The climate network is constructed by mapping the data to a spatial grid of $N$ approximately uniformly distributed points using the Fekete algorithm \citep{Bendito2007}. For visualization purposes, only every $10$th grid point is plotted. Dashed lines indicate the latitudinal range of the analysis $[20\degree{S}, 60\degree{N}]$.
        \textbf{b} Every grid point is associated with a precipitation time series. By thresholding locally at the 90th percentile ($Q_{0.9}$), we obtain binary event series ($Evs$).
        \textbf{c} The Event Synchronization algorithm \citep{Quiroga2002} is employed to evaluate the statistical dependencies between all pairs of time series (i.e. a total of $N\times N$ comparisons).
        The time series and event series in \textbf{b} and \textbf{c} are just for illustrative purposes.
        \textbf{d} The adjacency matrix $\mathbf{A}$ of the network characterizes the linkages between nodes, delineating the network's underlying topology. Each black dot represents a statistically significant synchronization between two nodes. Again, this matrix is for illustrative purposes.
        \textbf{e} Communities within the climate network represent groups of nodes characterized by stronger internal connectivity compared to their connections with nodes outside the community. The communities are determined by identifying blocks in the adjacency matrix $\mathbf{A}$ by using a probabilistic community detection algorithm.
        \textbf{f} Multiple runs of the community detection algorithm are performed to ensure the stability of the community pattern. The membership likelihood of each node to a community is determined by the number of times it is assigned to a specific community. Here it is plotted for the community comprising North India (NI) and the Sahel Zone (SZ). The hatched areas in \textbf{e} and \textbf{f} indicate regions with only few wet days, which are excluded from the analysis.
    }
    \label{fig:cd_sketch}
\end{figure}
The community detection approach is schematically visualized in Fig.~\ref{fig:cd_sketch}.
On a regular (rectangular) grid, grid points are spatially closer together the closer they are to the poles and therefore also more likely correlated. This could cause biases in the community detection algorithm later (see Sec.~\ref{sec:community_detection}). To avoid these confounding correlation effects, we first map the data to a grid of spatially approximately uniformly distributed points using nearest-neighbor interpolation (Fig.~\ref{fig:cd_sketch}~a) by employing the Fekete algorithm \citep{Bendito2007}.
For computational reasons, the distance between two points of the new grid is set to around $111$~km, which corresponds to the spatial distance between two points at the equator of a regular Gaussian $1^{\circ}$ grid, resulting in a total of approximately $9600$ grid points.
We only consider ``wet days'', i.e., days with rainfall sum of at least 1 mm. To capture the most intense rainfall events and analyze their characteristics, we determine ERE days per location as those where the daily precipitation sum exceeds the $0.9$th quantile ($Q_{0.9}(\cdot))$ of all wet days at that specific location (Fig.~\ref{fig:cd_sketch}~b).

We use climate networks to determine regions of significantly synchronously occurring EREs. The climate network is constructed as follows:
Let a spatiotemporal dataset of rainfall time series be denoted as $\mathbf{X}\in \mathbb{R}^{N\times T}$, where $N$ represents the spatial locations and $T$ is the number of time points.
The climate network is then defined as $\mathcal{G}=(V, E)$, where each geographical location $i \in 1,\dots, N$ corresponds to a node $n_i\in V$, and is associated with the rainfall time series $\vec{x}_i = x_{i,0},\dots,x_{i,T} \in \mathbf{X}$. $E$ represents the set of edges within the network. An edge between two nodes $n_i$ and $n_j$ encodes a robust statistical dependence between the time series $\vec{x}_i$ and $\vec{x}_j(t)$ labeled as edge $e_{ij}\in E$.

The Event Synchronization algorithm \citep{Quiroga2002} is employed to evaluate the statistical dependencies as a measure of the synchronization between all pairs of time series in the network Fig.~\ref{fig:cd_sketch}~c).
This method involves counting concurrently occurring events within event sequences from different locations, allowing a temporal gap -- the so-called dynamical delay $\tau$ -- between events in these sequences to account for time deviations of at most $10$ days (see SI Sec.~\ref{sec:climate_network_si}).
The synchronization strength between locations is assessed by summing up synchronous time points across all event pairs and their statistical significance is estimated using a null-model test.
The adjacency matrix $\mathbf{A}$ of a network characterizes the linkages between nodes, delineating the network's underlying topology.
It is a mathematical representation of these connections and captures the presence or absence of links between nodes expressed as a $N\times N$ matrix, where $A_{i,j}=1$  indicates that events at location $i$ are statistically significantly followed by events at location $j$, implying that we place an edge from node $n_i$ to $n_j$ and set $A_{i,j}=1$. For details, we refer the reader to Sec.~\ref{sec:climate_network_si} or to the detailed descriptions in \cite{Malik2010, Boers2014, Boers2019, Strnad2023}.

Communities within the climate network represent groups of nodes characterized by stronger internal connectivity compared to their connections with nodes outside the community. Keeping in mind that the network is constructed based on pairs of event series exhibiting significant synchronicity in the occurrence of EREs, these communities correspond to spatial regions that can span considerable geographical distances (Fig.\ref{fig:cd_sketch}~e) and where the occurrence of EREs in such cases is likely attributable to one (or even multiple) shared underlying physical mechanism(s) \citep{Strnad2023}.
Mathematically, communities are determined by identifying blocks in the adjacency matrix $\mathbf{A}$. Very common is an approach that uses a stochastic block model (SBM). The SBM serves as a generative model for random graphs, generating community structures - specifically, subsets of nodes referred to as ``blocks''. These exhibit distinct connectivity patterns, being  connected with one another at particular densities.
The model implementation used in this work is the network analysis package \textit{graph\_tool} \citep{peixoto_graph-tool_2014} favored for its computational efficiency and its probabilistic output.
Additionally, this implementation can identify the optimal number of communities in the data based on the principle of parsimony.
For more in-depth details, we refer the reader to \citet{Peixoto2014, Peixoto2019}.
As the community detection algorithm is inherently probabilistic, we repeat it $100$ times and use the distribution of different community detection outputs to quantify the membership likelihood of spatial locations (Fig.~\ref{fig:cd_sketch}~f). We define the membership likelihood as the number of times a node is assigned to a specific community. For more details on the community detection algorithm for climate networks, we refer the reader to \cite{Strnad2023}.

\subsection{Lagged Synchronous Rainfall Index} \label{sec:sync_ere_index}
We generalize the concept of the community-specific synchronous ERE index, denoted as $SRI(t)$, introduced by \cite{Strnad2023}, to assess for a community its degree of synchronization over time for a specified delay $\tau_{\text{lag}}$.
The index is computed for a particular set of locations $A$, describing one community within the network.
For each time step $t$, we count the number of EREs that occur in all event series $e_n$ in $A$ and are followed by events in all event series $e_m$ in $B$ at a lag $\tau_{\text{lag}}$:
\begin{equation}
    SRI_{\tau_{\text{lag}}}(t) = \sum_{n \in A}\sum_{n \in B} e_n(t)\cdot e_m(t-\tau_{\text{lag}})\; . \label{eq:sync_ere_index_basic}
\end{equation}
We further observe that the lag between synchronous events in $A$ and $B$ can vary. Thus, we also count all synchronizations between $A$ and $B$ within a certain range $[\tau_{\text{min}}, \tau_{\text{max}}]$, by using eq.~\ref{eq:sync_ere_index_basic}.
The synchronous rainfall index $SRI$ is estimated as:
\begin{equation}
    SRI(t) = \sum_{\tau = \tau_{\text{min}}}^{\tau_{\text{max}}} SRI_{\tau}(t)\; . \label{eq:sync_ere_index}
\end{equation}

This counting process enables us to quantify the frequency of EREs within a set of locations $A$ that are followed by events in a set of locations $B$ within a range of possible lags.
To further pinpoint the points in time of exceptionally strong (lagged) synchronization, we identify the local maxima in the time series $SRI$ that are above the 90th percentile and define these as the \textit{most synchronous days} (MSDs).

\subsection{Latent Space Clustering of Propagation Pathways} \label{sec:latent_space_clustering}

\begin{figure}[!htb]
    \centering
    \includegraphics[width=1.0\textwidth]{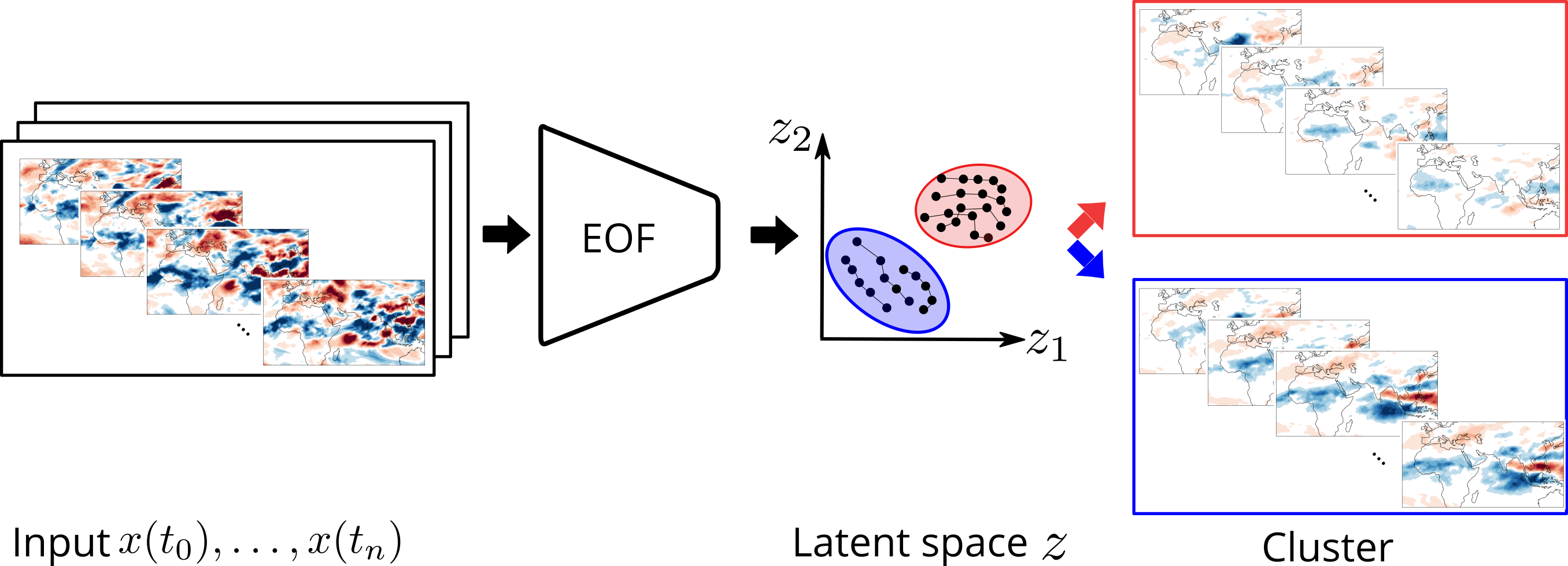}
    \caption{\textbf{Sketch of the latent space clustering.}
        The propagation pathways are clustered via traces in the latent space $z$. The high-dimensional input field is transformed to a low-dimensional latent space which is denoted by the encoder.
        In our case, the encoder function is based on a Multivariate Principal Component Analysis (MV-PCA) on 200-hPa $u$ and $v$, 500-hPa vertical velocity $\omega$ and 400-hPa relative humidity.
        Input fields, $x(t_1),\dots,x(t_n)$, for a propagation sample of $n$ consecutive time steps are encoded to the latent space $z$ (displayed as black dots). These build a trace in the latent space (visualized by connected lines). These traces are clustered by a clustering algorithm, the clusters are visualized as red and blue regions in the latent space.
        These correspond to the composite anomaly maps of the input field in the original space.
    }
    \label{fig:method_sketch}
\end{figure}

To obtain meaningful clusters of the propagation pathways, we propose a variation of latent space clustering, schematically visualized in Fig.~\ref{fig:method_sketch}, as it was applied in, e.g., \citep{Schloer2023}.
We initially employ empirical orthogonal function (EOF) analysis, also known as principal component analysis (PCA).
This well-established technique is used for reducing the dimensionality of spatiotemporal data, while retaining the largest possible fraction of its original variance.
A spatiotemporal field $x \in \mathbf{X}$ can consist of multiple variables. At time $t$ a transformation, $z_{t} = e(x_{t})$, is applied to map the high-dimensional space ,$\mathbb{R}^N$, to a lower-dimensional space,$\mathbb{R}^M$, where $M\ll N$.


To cluster the dynamics of the propagation up to a lag of $+15$ days, we compute the traces in the latent space $z$, where each sample $s$ concatenates the steps in the latent space $z$ from day $0$ to day $+15$: $s=\{z_{0},z_{1},\ldots,z_{15}\}$.
Therefore, each sample $t$ is a vector of length $15 \times n_{EOFs}$, where $n_{EOFs}$ is the number of EOFs used for the clustering. To select the EOFs that best represent the propagation pathways, we consider the EOFs whose PCs have the highest correlation coefficient with the lagged synchronous index $SRI(t)$ (eq.~\ref{eq:sync_ere_index}).

We employed the $k$-means clustering algorithm to cluster the traces in the latent space $z$. To determine the optimal number of clusters, we utilized the silhouette score \citep{Rousseeuw1987}.It measures the similarity of an object to its own cluster compared to other clusters, ranging from -1 to 1. A high value indicates a strong match to its cluster and a poor match to neighboring clusters. A high average silhouette score across all samples indicates an appropriate clustering configuration. We performed a grid search over the number of clusters and EOFs, ranging from $2$ to $10$ clusters and $1$ to $10$ EOFs, in order to identify the optimal clustering configuration (see Sec.~\ref{sec:grid_search_mvpca}).
Here, the best clustering is obtained for $2$ clusters and a single EOF (SI Fig.~\ref{fig:grid_search_mvpca}).

\subsection{Estimation of Conditional Probabilities} \label{sec:cond_indpendence_test}
The probability for the occurrence of synchronous rainfall days within a set of locations (denoted as $s=1$) under a condition, $a$, is calculated as follows.
Define the set of MSDs (i.e. days where we observe a high number of synchronizations between North India and the Sahel, see Sec.~\ref{sec:sync_ere_index}) as $S$ and the set of days that fulfill the condition $a$ as $A$.
Then
$$
    P(s=1|a) = \frac{P(s=1, a)}{P(a)} = \frac{||S \cap A||}{||A||}
$$
describes the conditional probability for synchronous events under condition $a$.
Here, $||\cdot||$ denotes the set cardinality and $S\cap A$ the intersection of $S$ and $A$.
Accordingly, the conditional probability for a second condition $b$ with a set of days $B$ is computed as:
\begin{equation}
    P(s=1|a, b) = \frac{P(s=1, a, b)}{P(a, b)} = \frac{||S \cap A \cap B||}{||A\cap B||} ~ .  \label{eq:cond_probs}
\end{equation}
A corresponding null model is estimated by counting the days of maximum synchronization $||S||$ divided by the total number of days. Therefore, by construction, the null model is $\leq 0.1$ (as our threshold for defining the MSDs was on the 0.9 quantile), and thus, the upper limit for the null model is $P_{\mathrm{null model}}(s=1) = 0.1$.

\section{Results}
\subsection{Lagged extreme rainfall synchronization between North India and the Sahel}
\begin{figure}[!tb]
    \centering
    \includegraphics[width=1\linewidth]{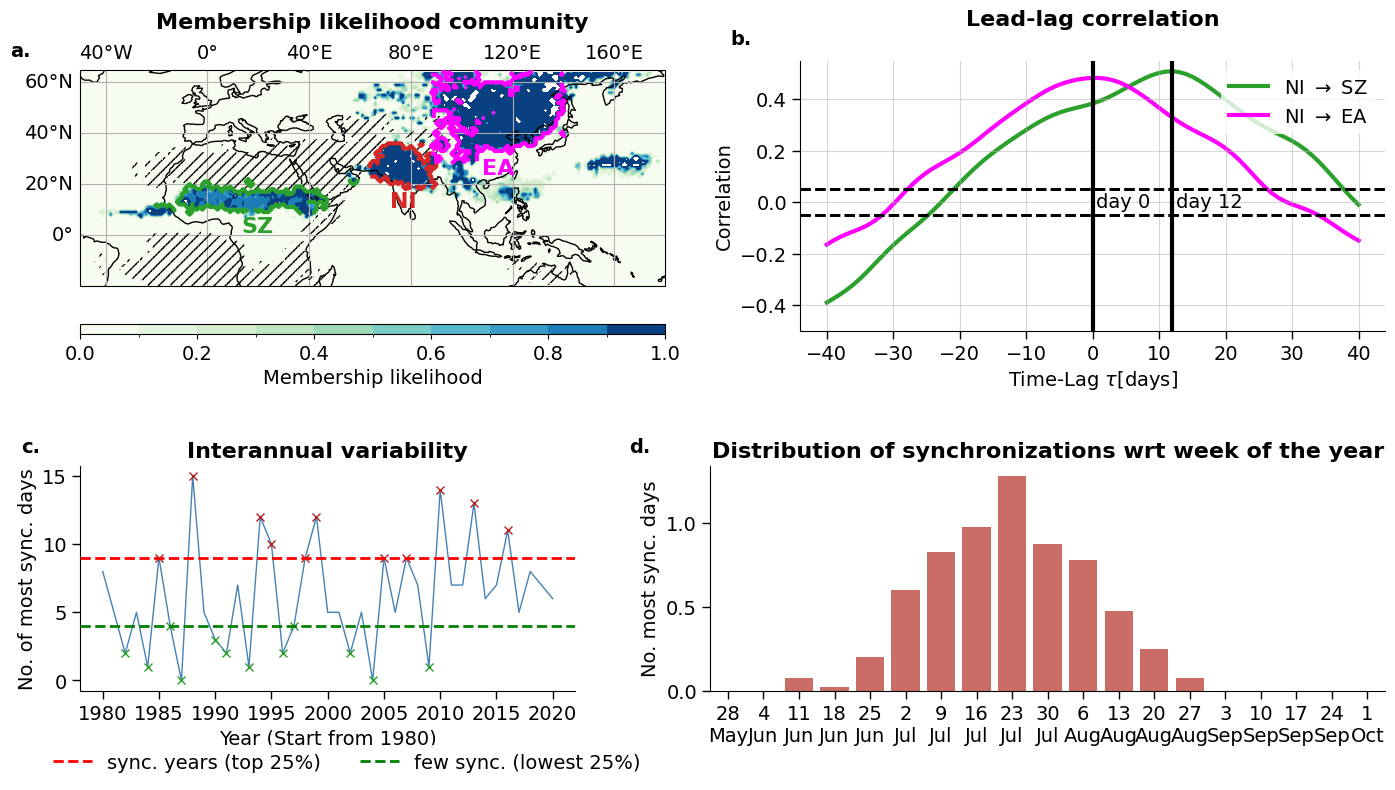}
    \caption{\textbf{Synchronization pattern of extreme rainfalls between North India and the Sahel}.
        \textbf{a} Regions of statistically significant synchronization between spatial locations are plotted according to their membership likelihood of being in the community comprising NI (red contour) and SZ (green contour).
        The communities are determined using a probabilistic hierarchical community detection algorithm based on the Stochastic Block Model and overlaps of $100$ independent runs.
        Hatched areas indicate regions with few wet days, which are excluded from the analysis.
        We find a community that comprises the regions of North India (NI), East Asia (EA), and the Sahel Zone (SZ). The connection between NI (red contours), SZ (green contours), and EA (magenta contours) is investigated.
        \textbf{b} Lead-lag correlations in steps of days between time series obtained from counting the number of extreme events of locations in the NI and the SZ (green line) and NI and EA (magenta line). The dashed line indicates the 95\% confidence interval for a statistically significant correlation.
        The maximum correlation occurs at a lag of $+12$ days (vertical solid line), i.e. EREs in NI are \textit{followed} by EREs in SZ by around \textit{12 days}, while the EREs in NI and EA typically occur synchronously.
        \textbf{c} Interannual variability of the synchronization between NI and SZ is investigated. The most synchronous days (MSDs) are counted per year. The 75th (25th) percentile of years with the highest (lowest) number of days identified as MSD are labeled as synchronous (dry) years, marked by red (green) markers.
        \textbf{d} For the MSDs, we estimate the distribution over the week of the year in the observation period from June to September.
    }
    \label{fig:basic_pattern}
\end{figure}

\paragraph{Pattern of the synchronization}
We identify a distinct synchronization pattern encompassing North Asia; in particular, North China, North India (NI) including the Indo-Gangetic Plain, the Himalayan foothills, delineated by the Himalayan mountain chain, and the Sahel Zone (SZ) bounded between the tropical rainforest and the Sahara desert (Fig.~\ref{fig:basic_pattern}~a).
Our investigation focuses on exploring why the EREs manifest synchronously (within a certain time range) over this large spatial region.
The regions are part of a ``community'' within the climate network constructed by estimating patterns of simultaneously occurring extreme rainfall events of a domain restricted to the tropics and boreal subtropics (see Sec.~\ref{sec:community_detection}).
The high spatial coherence indicates that the community is a stable manifestation of the synchronization pattern associated with synchronous EREs (Fig.~\ref{fig:msl_all}).
It can be partly attributed to the Eurasian Wave Train associated with the Silk Road pattern \citep{Gupta2022}.
In the following, we thus focus on the synchronization between North India (red contour in Fig.~\ref{fig:basic_pattern}~a) and the Sahel Zone (green contour in Fig.~\ref{fig:basic_pattern}a).

\paragraph{Time lags of the synchronization} Synchronizations were observed to peak in strength at a time lag of around $12$ days, on average, throughout boreal summer (June, July, August, September) (Fig.~\ref{fig:basic_pattern}~b). This typical time delay between EREs within the NI region and the SZ locations is estimated by lead-lag correlation analysis.
The number of occurrences of synchronization patterns between NI and SZ is not constant but varies substantially between different years (Fig.~\ref{fig:basic_pattern}~c).
We obtain this yearly fluctuation by counting the number of MSDs per year.
For later reference, we define the top $25~\%$ as the synchronous years (red markers in Fig.~\ref{fig:basic_pattern}~c) and the lowest $25~\%$ as the least synchronous years (green markers in Fig.~\ref{fig:basic_pattern}~c).
We also observe that the synchronization occurs mainly during the core monsoon season from July through mid-August (Fig.~\ref{fig:basic_pattern}~d). The distribution over the JJAS period is estimated by the week of the year for all identified MSDs.
Therefore, hereafter we focus on synchronization events in July and August, which we consider to be the background state under which most synchronization happens.
While the occurrence of synchronous events is high mainly during July and August, coinciding with the period of highest daily rainfall sums in India, there is still strong interannual variability within these months.

\paragraph{Anomalous convection in the northwest of India}

\begin{figure}[!tb]
    \centering
    \includegraphics[width=1.\linewidth]{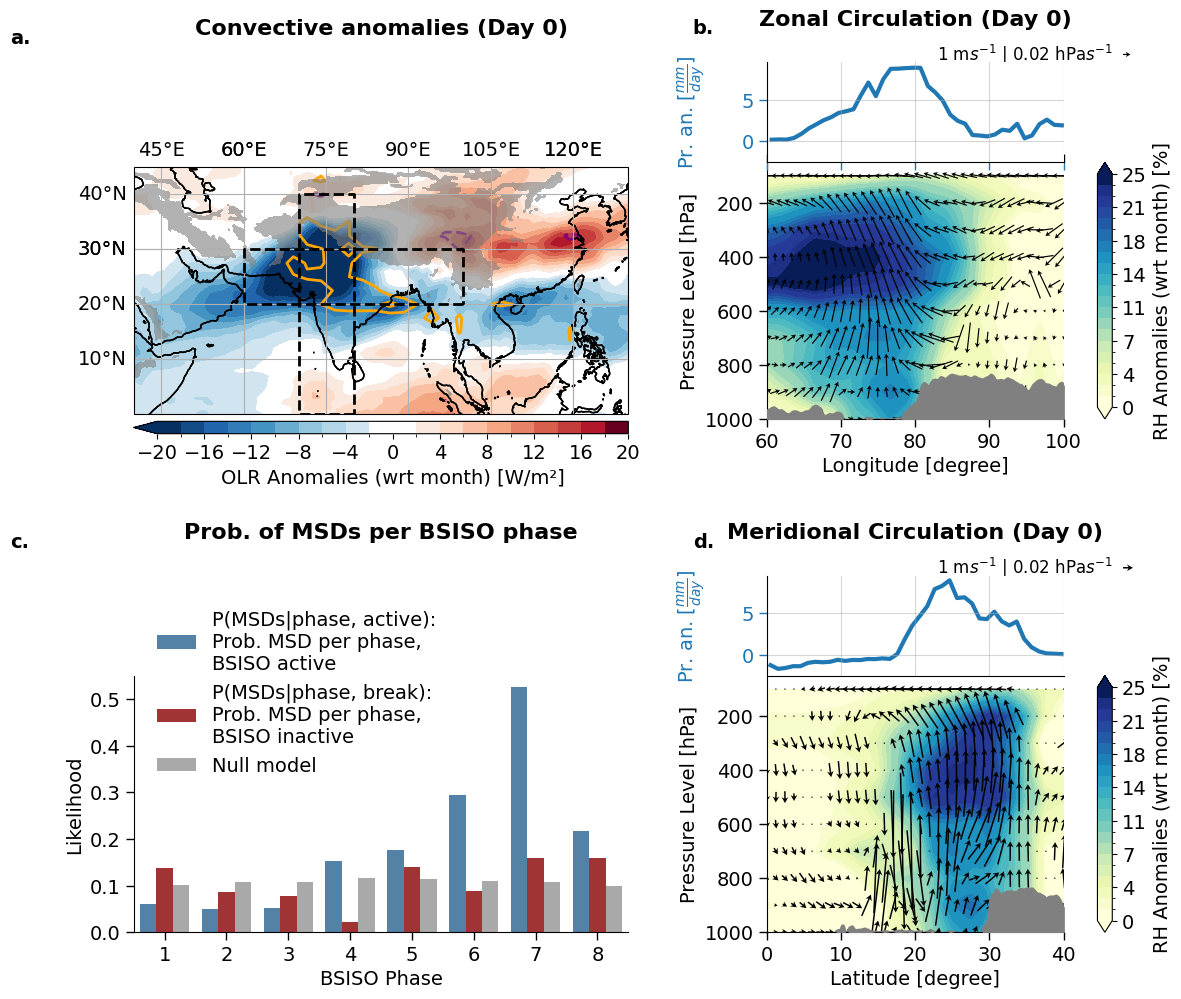}
    \caption{\textbf{Convective anomalies in North India.}
        Panel \textbf{a} shows composites of the MSDs for the Outgoing Longwave Radiation (OLR) superimposed by the vertical velocities $\omega$ at $400~$hPa  (measured in Pa/s) by colored contours. Orange solid (purple dashed) contours denote anomalously rising conditions for regions with anomalies larger (smaller) than $3\,$Pa/s. Grey shading represents the Himalayan mountains.
        Panel \textbf{b} (\textbf{d}) visualizes the zonal (meridional) circulation averaged between $20\degree{N}-30\degree{N}$ ($70\degree{E}-80\degree{E}$), visualized by dashed rectangles in \textbf{a}.
        All composites are computed for the conditions on Day 0.
        Panels \textbf{b},\textbf{d} are split into two parts: on top the meridionally (zonally) averaged precipitation anomalies are displayed; on the bottom, the vertical circulation is shown by composites of relative humidity anomalies (shading) and wind fields (arrows). In \textbf{a}, \textbf{b}, \textbf{d} colored contours denote statistical significance at a $95\%$ confidence level using a two-sided $t$-test. Grey contours show the highest $10\%$ orography.
        The wind fields in the zonal (meridional) circulation plots are estimated using the meridionally (zonally) averaged $u$ ($v$) anomalies, measured in m~s$^{-1}$, and the vertical velocity $\omega$ in the horizontal direction, measured in hPa~s$^{-1}$. Only statistically significant arrows at a $95\%$ confidence level using a two-sided $t$-test are shown.
        Panel \textbf{c} shows the condition probabilities (Sec.~\ref{sec:cond_indpendence_test}) of the most synchronous days (MSDs) conditioned on active (inactive) BSISO phases marked by blue (red) bars. The grey bars denote the likelihood of a respective null model that randomly distributes the MSDs over the BSISO phases with respect to the relative occurrence of each phase.
    }
    \label{fig:NI_convection}
\end{figure}
The synchronization is initiated by anomalously strong convection in northwest India (Fig.~\ref{fig:NI_convection}~a).
We observe two characteristics.
Firstly, the anomalously strong convection (Fig.~\ref{fig:NI_convection}) in the northwest also coincides with the locations where the strongest rainfall occurs (Fig.~\ref{fig:NI_convection}~b,d) as is usually expected in the tropics. Secondly, we also note a deepening of the monsoon trough for the band of enhanced negative OLR anomalies at around $15\degree{N}$ (Fig.~\ref{fig:NI_convection}~a).
We further observe a strong anomalous moistening of the upper atmosphere in Fig.~\ref{fig:NI_convection}~b and d, as well as anomalous easterly zonal winds in the upper atmosphere around $200~$hPa.

The anomalously strong convection in northwest India already suggests the presence of the BSISO.
An analysis based on conditional probabilities (see Sec.~\ref{sec:cond_indpendence_test}) supports this impression and shows a substantially increased likelihood for MSDs when the BSISO is in phases 6 and 7 (Fig.~\ref{fig:NI_convection}~c, SI Fig.~\ref{fig:bsiso_phases_all_tps}~e).
The convective band is thus likely to be associated with the active phase of the BSISO, which is characterized by heavy rainfall (see e.g., \cite{Kikuchi2012, Kiladis2014, Kikuchi2021}).

Further, during the MSDs for both clusters, we observe an increased inflow of moist air from the Arabian Sea via the cross-equatorial Somali Jet (Fig.~\ref{fig:ivf_flow}~a, b). The Somali Jet is a major source of moisture during the ISM \citep{Rai2018}.
It is therefore consistent that the synchronization pattern is mainly established during the core monsoon season (Fig.~\ref{fig:basic_pattern}~c) when the Somali Jet is most active.
For dry years (green markers in Fig.~\ref{fig:basic_pattern}~c), the synchronization is less pronounced, and the enhanced Somali-Jet inflow vanishes (Fig.~\ref{fig:ivf_flow}~c), indicating that the moisture flux from the Arabian Sea is crucial for the initiation of the synchronization.
When this moisture flux hits the western Himalayan mountains, it leads to forced deep convection (Fig.~\ref{fig:NI_convection}~a, b, d). The convection that initiates the synchronization is thus -- at least partially -- forced by the orography.

\subsection{Propagation pathways of the synchronization}
\begin{figure}[!tb]
    \centering
    \includegraphics[width=1.\linewidth]{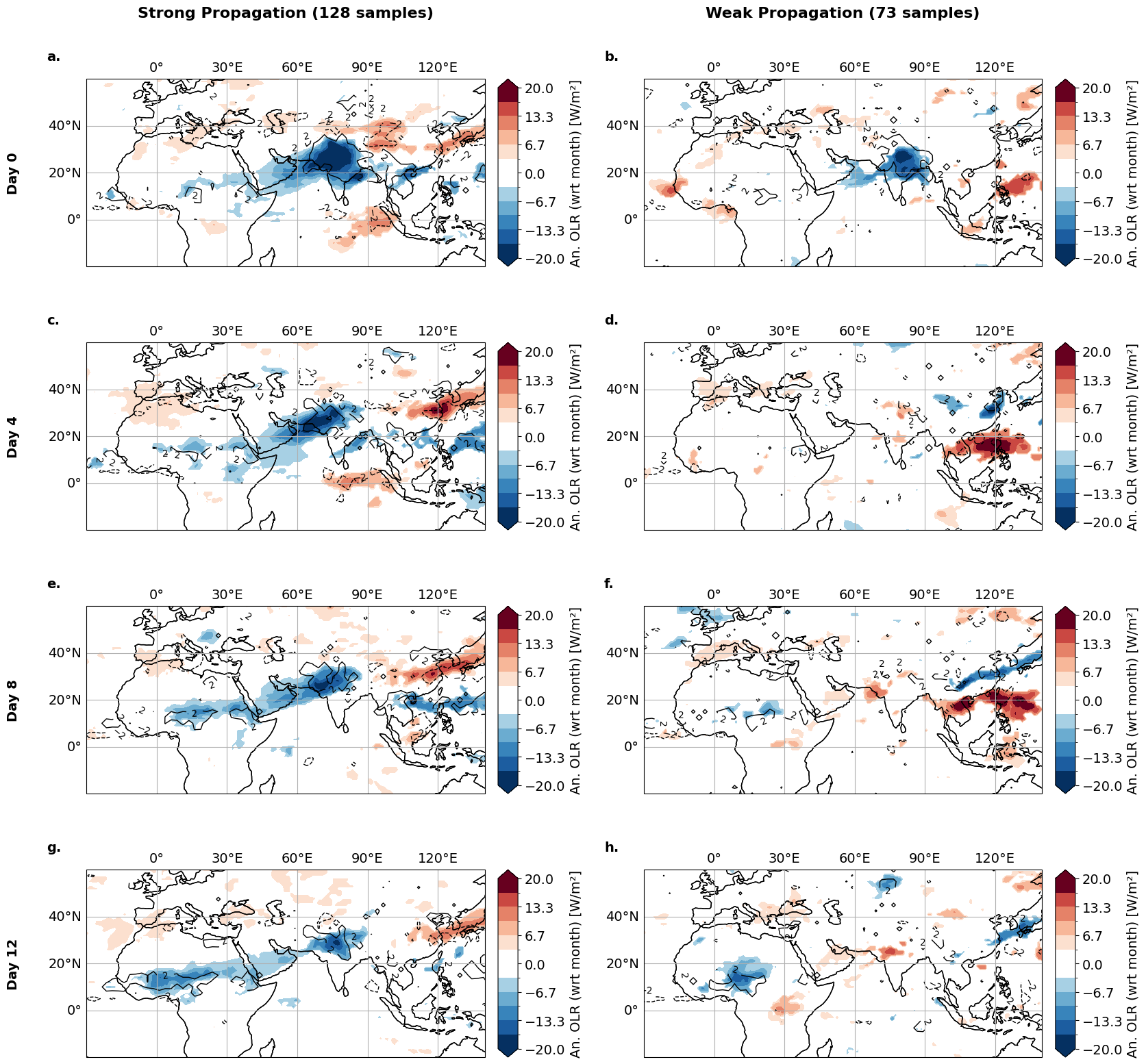}
    \caption{\textbf{Propagation for OLR and vertical velocity.}
        We cluster the propagation pathways based on the most synchronous days using the lagged synchronous index (compare Fig.~\ref{fig:basic_pattern}~b).
        The first column (panels \textbf{a,c,e,f}) shows the first cluster of propagation pathways using OLR. The second column (panels \textbf{b,d,f,h}) shows the second cluster.
        We create composite anomaly maps (estimated with respect to the month of the year) for days 0 to +12 after initialization of outgoing longwave radiation (OLR) in shading, overlapped by vertical velocity $\omega$ at $400\,$hPa (measured in Pa/s) in black line contours. Black solid (dashed) contours denote anomalously rising (sinking) air with $\omega$ anomalies larger (smaller) than $2\,$Pa/s.
        For visual reasons, composites are only shown in steps of $+4$ days, starting from day $0$.
        Colored areas denote statistically significant values at $95~\%$ confidence level.
    }
    \label{fig:propagation_olr}
\end{figure}

The synchronization between North India and the Sahel Zone is not instantaneous but lags by almost two weeks (Fig.~\ref{fig:basic_pattern}~b). However, when investigating individual years, we observe that the mean lag is not constant but varies across years (SI Fig.~\ref{fig:yearly_lead_lag}).
We thus investigate potential variations in the diversity of the synchronization propagation pathways.
From the observation of the convective initiation and the vertical circulation plots (Fig.~\ref{fig:NI_convection}), we expect that the propagation pathways are somehow driven by the convective activity, the subsequent moistening of the upper atmosphere and anomalous processes in the upper troposphere (around $200~$hPa).
This motivates the choice of the following variables for a multivariate PCA (MV-PCA).
We use vertical velocity $\omega$ at $500~$hPa as a proxy for convective activity, relative humidity at $400$~hPa to account for the moistening of the mid and upper troposphere and horizontal ($u$) and horizontal ($v$) winds at $200~$hPa for the propagation. Where clusters of consecutive days occur in the set of MSDs, we remove all but the first to avoid including the same event more than once in our composite. The MV-PCA helps to substantially reduce the dimensionality of the fields and to identify the PCs of the OLR field that are associated with the synchronization (see Sec.~\ref{sec:latent_space_clustering} for details).

We create a trace in the latent space of $15$ consecutive days for each time point that is identified as an MSD.
We use these traces as input features for a $k$-means clustering algorithm.
The clustering with the highest silhouette score is obtained for $k=2$ clusters and $1$ EOF (Fig.~\ref{fig:propagation_olr}).
We thus find two distinct clusters, and consistently, the years in which these clusters occur are largely distinct from each other in terms of timing (Fig.~\ref{fig:sst_background}~d).
We observe that the propagation of the negative OLR anomalies is accompanied by a strong rising motion of the air in the mid-troposphere (Fig.~\ref{fig:propagation_olr}~a-h).
Composite anomaly maps are computed for each cluster and each day in steps of $+3$ days (Fig.~\ref{fig:propagation_olr}) revealing two different propagation clusters:
\begin{itemize}
    \item \textbf{Strong Propagation} The first column in Fig.~\ref{fig:propagation_olr}, demonstrates the propagation of anomalous OLR in steps of 4 days, from day 0 to day 12. Note that negative OLR anomalies often coincide with intense rainfall.
          We see a region of negative OLR anomalies propagating from North India (NI) toward the Sahel Zone (SZ). These are largely confined to NI and re-emerge later in the SZ. Strongly anomalous OLR values over the full range of the SZ are observed from around day $+8$ on (Fig.~\ref{fig:propagation_olr}~e). The associated rainfall lags the rainfall in North India by around $10-12$ days.
    \item \textbf{Weak Propagation} The second column of Fig.~\ref{fig:propagation_olr} does not show a clear propagation pattern and the anomalies over SZ are of reduced intensity.
          Substantially different from the first cluster is not only the intensity but also the speed of the propagation. This is reflected by strongly anomalous OLR values are observed from around day $+10$ on (Fig.~\ref{fig:propagation_olr}~g). The rainfall here lags the rainfall in North India by around $+12-15$ days.
\end{itemize}
The clusters also exhibit different background conditions in the Indian Ocean and the Western Pacific.
While the first cluster shows persistent convective anomalies over NI and parts of the Maritime Continent, the second cluster reveals a developing pattern of reduced convection at NI, parts of the Arabian Sea and around the equatorial Western Pacific (Fig.~\ref{fig:propagation_olr}~h--l).
However, as the first cluster shows a more intense propagation and the number of samples in cluster 0 is twice the number of samples in cluster 1, we suggest that the first cluster is the main manifestation of the mechanism resulting in the synchronization between NI and SZ, while the second cluster is a weaker variation of the main mechanism. Further, given the smaller cluster size, we cannot rule out that some reasonable fraction of events are not causally connected but arise by chance. Therefore, in the following, we will focus on the first cluster.

\subsection{Background conditions favoring the ISM--WAM synchronization}
\begin{figure}[!htb]
    \centering
    \includegraphics[width=1.\linewidth]{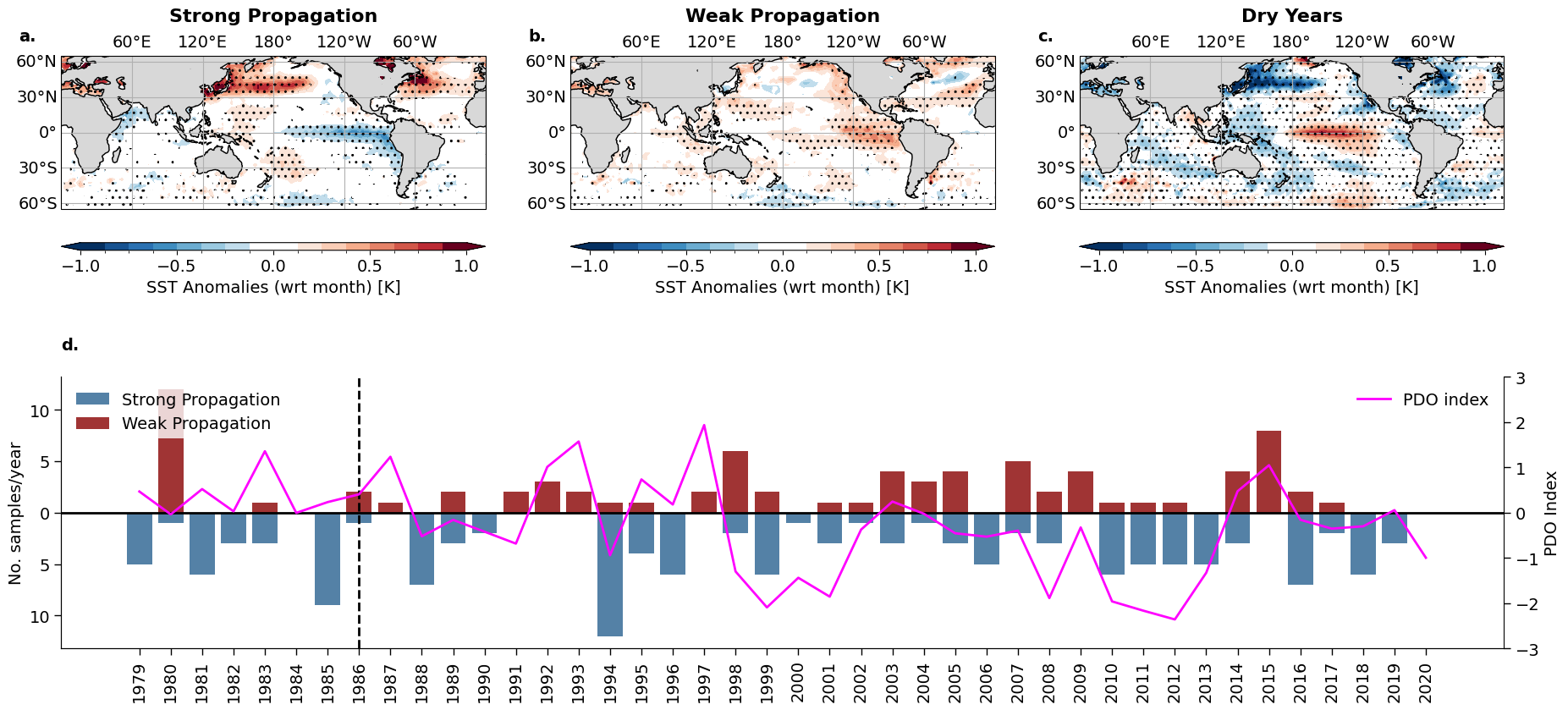}
    \caption{\textbf{Background SST states for latent cluster.}
        \textbf{a}--\textbf{c} shows the sea surface temperature (SST) background state conditions for the Strong Propagation (\textbf{a}) and Weak Propagation (\textbf{b}).
        For comparison, conditions for years with few/no synchronizations are also shown (\textbf{c}).
        Dotted areas indicate statistical significance at $95~\%$ confidence level using Student's $t$-test.
        Panel \textbf{d} shows the distribution of days classified as Strong and Weak Propagation alongside the PDO index (magenta line). The dotted vertical line indicates the year 1986 after which the clustered propagation pathways match the PDO phases quite well.
    }
    \label{fig:sst_background}
\end{figure}

\paragraph{SST background state enhancing synchronization likelihood}
We find that the background SST state shows distinct patterns for the two clusters.
The Strong Propagation cluster shows an anomalous cooling in the central Pacific (Fig.~\ref{fig:sst_background}~a) -- a La Ni\~na like pattern -- and a band of anomalously warm SSTs in the northwest Pacific at around $40\degree{N}$, which is known as the Kuroshio-Oyashio extension \citep{DiLorenzo2023, Joh2023}.
Together, these patterns in the Pacific Ocean resemble the patterns of a negative phase of the Pacific Decadal Oscillation (PDO) in JJAS (see Fig.~\ref{fig:PDO_phases}).
We find that starting from around 1986, the occurrences of samples that are clustered as Strong Propagation (Weak Propagation) are well aligned with the negative (positive) phase of the PDO (Fig.~\ref{fig:sst_background}~d) even though the match is not perfect.
The negative phase of the PDO favors the occurrence of La Ni\~na events \citep{DiLorenzo2023}.
La Ni\~na-like conditions lead to an anomalously wet monsoon season in North India through modulating the Walker circulation \citep{Xavier2007} and by assisting the northward propagation of the BSISO towards the north of India \citep{Strnad2023}.
We additionally observe associated cool SSTs over the Arabian Sea and the Bay of Bengal, which are probably due to the enhanced convection over the region, i.e. cooling the surface through increased insolation and precipitation. However, this might also be associated with a negative Indian Ocean Dipole which is often dynamically linked to La Ni\~na events \citep{Meyers2007, Cherchi2013}.

The second cluster does not show a strong SST pattern in the tropical Pacific Ocean (Fig.~\ref{fig:sst_background}~b).
We identify a weak warming pattern in the Eastern Pacific represented by anomalously warm SSTs in the Pacific Ocean (Fig.~\ref{fig:sst_background}~b).
However, this pattern is less pronounced than the La Ni\~na-like one, suggesting that the La Ni\~na background state provides an important intensification of the synchronization, even if the synchronization can also occur without a La Ni\~na event.
For years with only a few synchronizations, which we call dry years, we identify a more pronounced El Ni\~no-like pattern represented by anomalously warm SSTs in the central equatorial Pacific Ocean (Fig.~\ref{fig:sst_background}~c).
This is consistent with the observation that central Pacific El Ni\~no events are more often associated with a weakening of the Indian monsoon \citep{Fan2017} and thus also weakening the synchronization between North India and the Sahel.

\paragraph{Synchronization is guided by the Tropical Easterly Jet}
\begin{figure}[!htb]
    \centering
    \includegraphics[width=1.\linewidth]{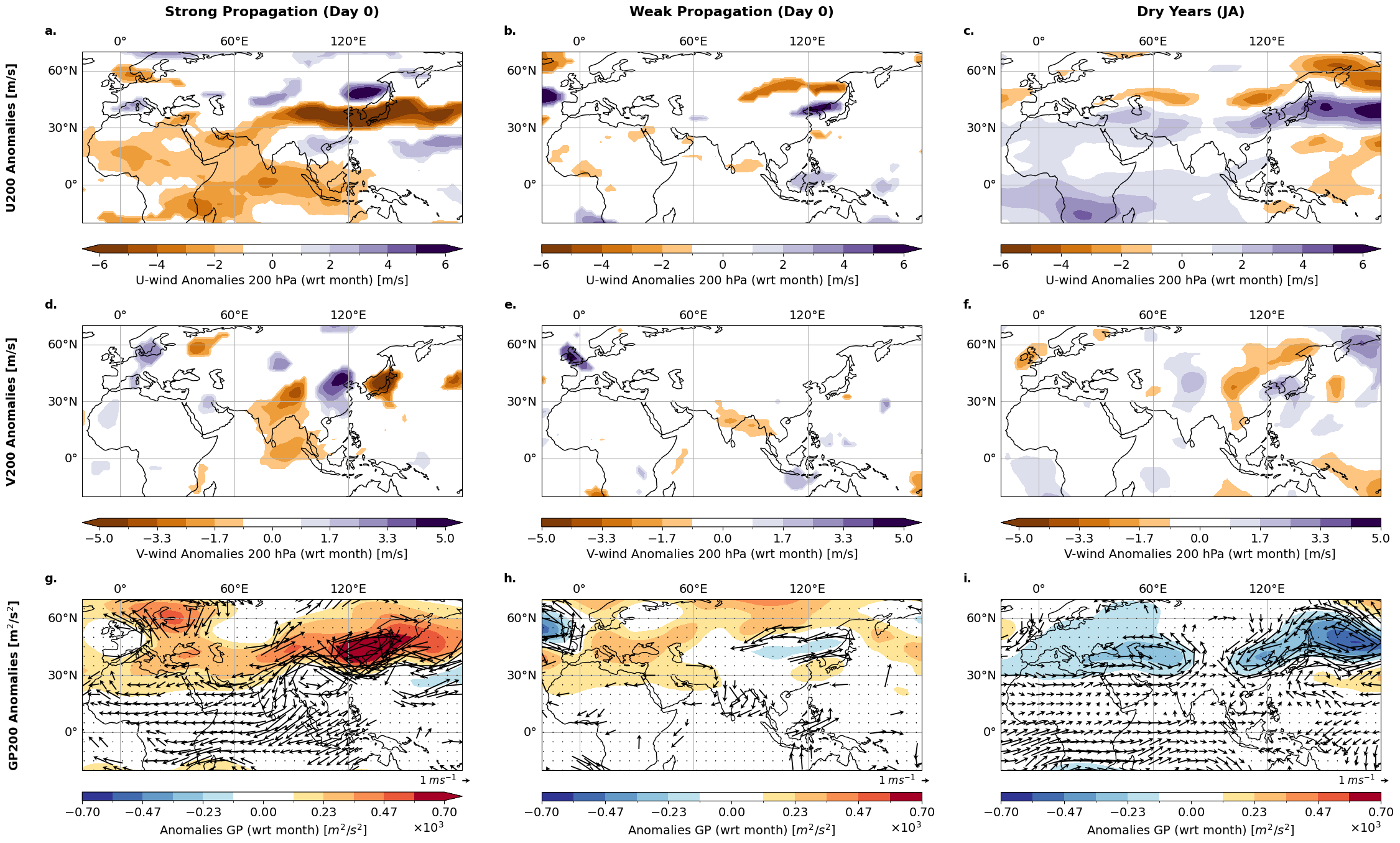}
    \caption{\textbf{Background states for propagation clusters.}
        We visualize different conditions for the samples of the days classified as most synchronous days that were clustered by multivariate PCA (Fig.~\ref{fig:propagation_olr}) into Strong Propagation (first column) and Weak Propagation (second column). For comparison, conditions for years with few/no synchronizations (see Fig.~\ref{fig:basic_pattern}~c) are also shown (third column).
        The first row (\textbf{a}--\textbf{c}) shows the composited zonal wind anomalies at 200~hPa.
        The second row (\textbf{g}--\textbf{i}) is as the first row but for meridional wind anomalies at 200~hPa.
        The third row (\textbf{j}--\textbf{l}) is also as the first row but for geopotential height (GPH) anomalies with arrows representing the horizontal wind anomalies at 200~hPa.
        All composited anomalies are computed with respect to the month of the year.
        In all panels, colored areas imply statistical significance at $95~\%$ confidence level using the Student's $t$-test and only statistically significant wind arrows are shown.
    }
    \label{fig:wind_background}
\end{figure}

The composite wind field at $200$~hPa for the Strong Propagation cluster shows an established corridor that zonally connects the area around the Yellow River Basin with North India and SZ (Fig.~\ref{fig:wind_background}~a). The flow is guided by the orography of the Tibetan Plateau and the pattern is especially prominent to the south of the Tibetan Plateau.
These patterns resemble the characteristics of an enhanced TEJ (Fig.~\ref{fig:def_tej_pattern}). The enhancement is likely a result of the La Ni\~na-like conditions (Fig.~\ref{fig:sst_background}~a) and the associated Walker circulation response.
Induced by La Ni\~na, there is anomalous heating over the Tibetan Plateau \citep{Duan2012}. The subsequent anomalous meridional temperature gradient leads to a strengthening of the TEJ core position over the Indian Ocean \citep[Fig~\ref{fig:wind_background}~a; also][]{Nithya2017}.
Further evidence is provided by the spatial correlation patterns obtained by correlating the 200-hPa zonal wind, 200-hPa meridional wind, and 500-hPa vertical velocity respectively with the lagged synchronous index (see eq.\ref{eq:sync_ere_index_basic}).
The correlation pattern resembles the structure of the tropical easterly jet \citep{Nicholson2021} (see also SI Fig.~\ref{fig:def_tej_pattern}).
The pattern of an enhanced TEJ is not visible for the Weak Propagation cluster (Fig.~\ref{fig:wind_background}~b). Here, the zonal winds are only moderately enhanced over the equatorial Indian Ocean.
In dry years, the zonal wind field is not enhanced over the Indian Ocean and the Bay of Bengal (Fig.~\ref{fig:wind_background}~c). On the contrary, the pattern resembles a weakened TEJ (Fig.~\ref{fig:def_tej_pattern}~b), which can be explained by the El Ni\~no-like SST pattern (Fig.~\ref{fig:sst_background}~c).

The 200-hPa meridional wind anomalies for composited time points of the Strong Propagation cluster (Fig.~\ref{fig:wind_background}~d) show a large-scale wave train pattern (Fig.~\ref{fig:wind_background}~g) ranging from Japan to North India.
The ridge in Fig.~\ref{fig:wind_background}~d is guided by an anticyclone-cyclone-anticyclone (A-C-A) circulation pattern. We observe these as a subtropical anticyclone near Japan (Fig.~\ref{fig:wind_background}~j) occurring together with an anomalous high known as the Bonin high, a cyclone-like structure over the South of China and anticyclone over the Himalayans, suggesting a relationship with the Silk Road Pattern \citep{Enomoto2003}.
The positive CGT drives southwesterlies into the northwest and north of India (Fig.~\ref{fig:wind_background}~g), which brings additional moisture from the Arabian Sea. At the local-synoptic level, this is similar to the Western disturbances (WD)-tropical storm interactions \citep{Hunt2021}.
This wave train pattern reduces to weak northerly anomalies in the second cluster (Fig.~\ref{fig:wind_background}~e) and vanishes altogether for dry years (Fig.~\ref{fig:wind_background}~f).

\paragraph{Combination of processes initiating the synchronization}
\begin{figure}[!tb]
    \centering
    \includegraphics[width=.9\linewidth]{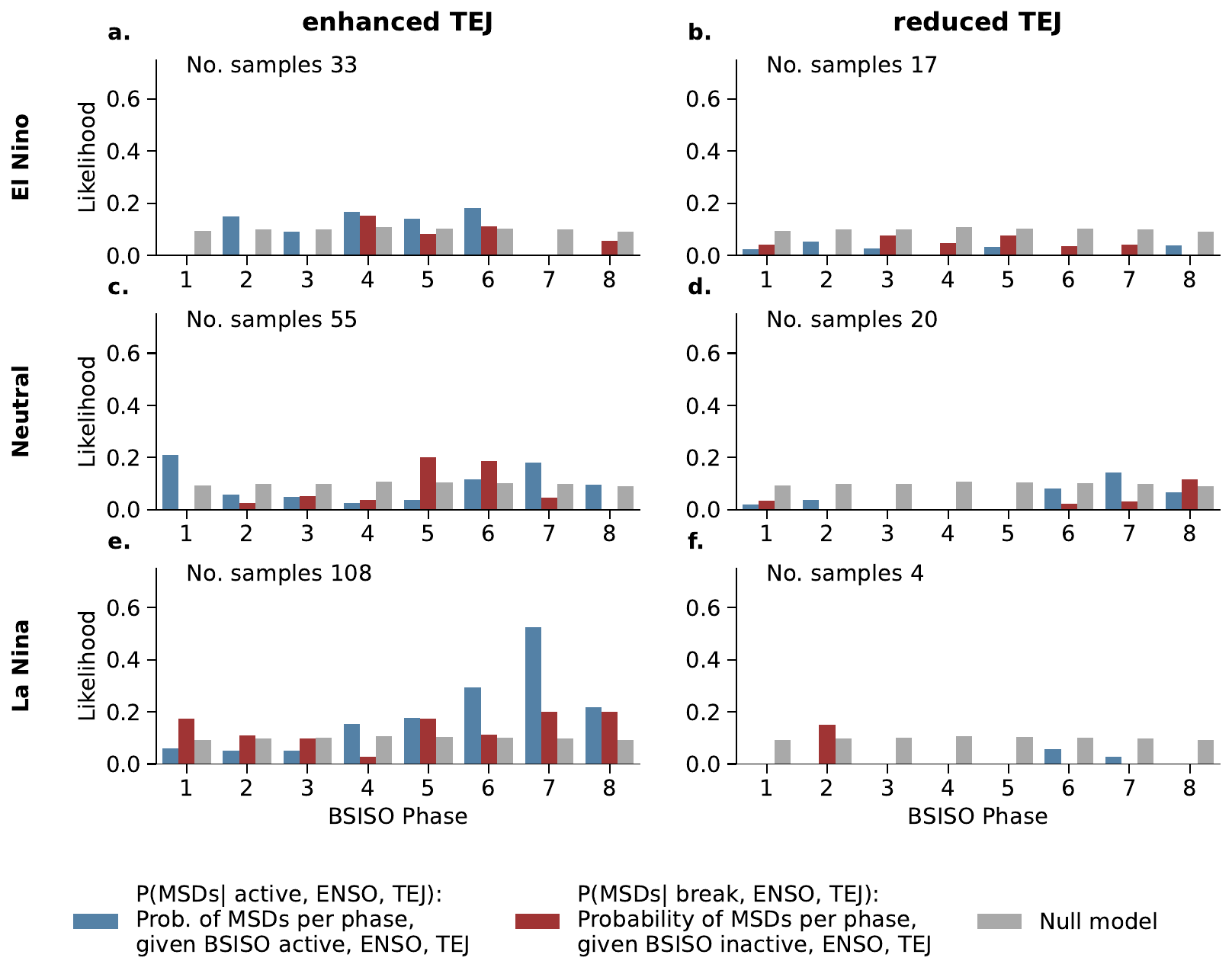}
    \caption{\textbf{Probabilities for the occurrence of synchronization conditioned on BSISO phases, ENSO and TEJ.}
        \textbf{a-f} Most synchronous days (MSDs) are defined using the $0.9$ quantile to ensure a robust sample size of possible events (Fig.~\ref{fig:basic_pattern}~b) when using the lagged synchronous index (compare Fig.~\ref{fig:basic_pattern}~d). Their probability of occurrence is conditioned on the state of the Boreal Summer Intraseasonal Oscillation (BSISO), the Tropical easterly jet (TEJ), and the respective phase of the El Ni\~no Southern Oscillation (ENSO) for the respective time points computed as in eq:~\ref{eq:cond_probs}.
        Blue (red) bars indicate an active (inactive) BSISO state conditioned on the state of ENSO and the TEJ. Grey bars denote the respective null model, i.e. MSDs are assigned randomly to days in the JJAS season and assigned to the distribution of BSISO phases.
    }
    \label{fig:likelihoods_bsiso_enso_tej}
\end{figure}
Taken together, the initiation of the synchronization mechanism can be attributed to a threefold process:
Firstly, a negative PDO phase favors La Ni\~na conditions (Fig.\ref{fig:sst_background}a), promoting deep convection over peninsular India, facilitated by the anomalous Walker circulation response (Fig.\ref{fig:NI_convection}b).
Secondly, this response supports the progression of the BSISO towards northwest India (Fig.~\ref{fig:NI_convection}~a), substantially enhancing the frequency of deep convection there. Thirdly, La Ni\~na-like conditions also enhance the TEJ (Fig.~\ref{fig:wind_background}~d, SI Fig.~\ref{fig:tej_correlation}).
We can provide strong statistical evidence for the importance of these three factors coming together. By estimating conditional probabilities (eq.~\ref{eq:cond_probs}) for the occurrence of MSDs, conditioned on the three impacting factors BSISO, ENSO and the TEJ (see Sec.~\ref{sec:data}), we observe that the conditional probabilities for a day in JJAS being a MSD conditioned on La Ni\~na, TEJ, as well as an active BSISO phase is substantially enhanced (Fig.~\ref{fig:likelihoods_bsiso_enso_tej}).
For a reduced TEJ we observe hardly any synchronizations, even if the BSISO is active and La Ni\~na conditions are present (Fig.~\ref{fig:likelihoods_bsiso_enso_tej}~b,d,f) emphasizing the necessary rolethe TEJ play in the propagation.

This is further supported by the observation that the pattern of anomalously negative OLR in Fig.~\ref{fig:NI_convection}~a occurs further north than in the full composites for BSISO phases 6 and 7 (SI Fig.~\ref{fig:bsiso_phases_all_tps}), whereas the subset of points in La Ni\~na years together with enhanced TEJ in BSISO phases 6 and 7 (Fig.~\ref{fig:bsiso_phases_LN_tej}~e,f) shows greater similarity to the composite MSD map (Fig.~\ref{fig:NI_convection}~a).

The combination of these factors results in an increased presence of upper-level moisture and a stronger TEJ.
The enhanced BSISO pushes moisture further north and northwest towards Pakistan.
In this region, the dry static stability is very low, and anomalous moisture readily triggers convection.
In consequence, we observe a significantly increased moistening of the middle to upper atmosphere from $500~$hPa to $200~$hPa (Fig.~\ref{fig:NI_convection}~b).

\subsection{Mechanism driving precipitation in the Sahel Zone}
\begin{figure}[!tb]
    \centering
    \includegraphics[width=1.\linewidth]{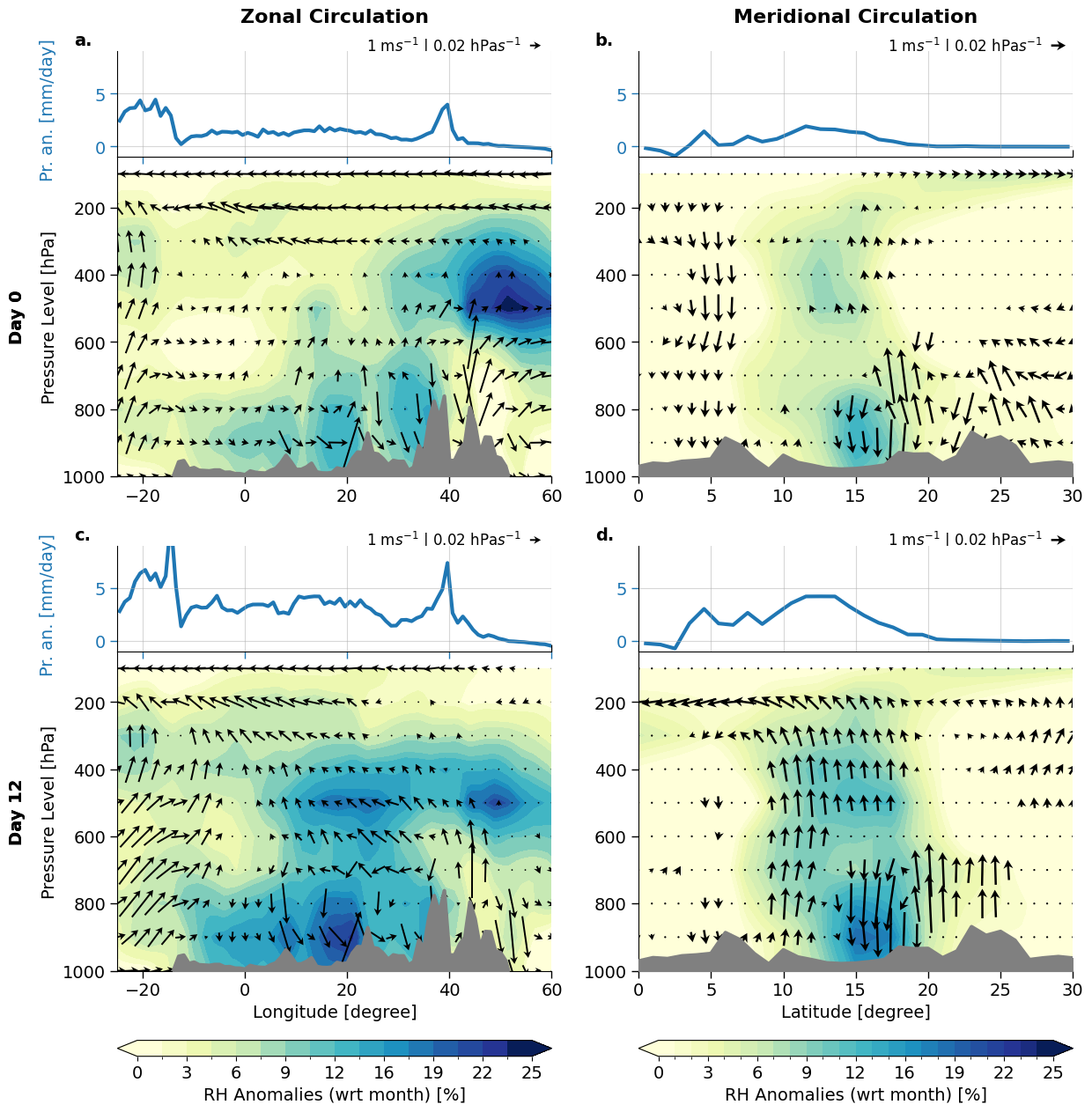}
    \caption{\textbf{Composite thermodynamic structure for synchronized EREs over the Sahel.} Conditions for the MSDs of the Strong propagation cluster are shown. The first column shows the zonal circulation averaged over $10\degree{N}-18\degree{N}$, while the second column shows the meridional circulation, averaged over $10\degree{E}-20\degree{E}$. The first row shows the conditions on Day 0 and the second row shows the average for days 10--12, which are the days of the strongest synchronizations (see Fig.~\ref{fig:msl_all}~b).
        Each subplot consists of two panels. The top panel shows the meridionally (zonally) averaged precipitation anomalies (with respect to the month). The bottom panel shows the vertical structure of the circulation in arrows and the filled contours show relative humidity anomalies.
        Grey contours show the orography. The wind fields in the zonal (meridional) circulation plots are estimated using the meridionally (zonally) averaged $u$ ($v$) anomalies, measured in m~s$^{-1}$, and the vertical velocity $\omega$ in the horizontal direction, measured in Pa~s$^{-1}$. Only statistically significant wind arrows at a $95\,\%$ confidence level using a two-sided $t$-test are shown.}
    \label{fig:SZ_rainfall}
\end{figure}
Our results indicate that the anomalous deep convection over NI (Fig.~\ref{fig:NI_convection}~b,c) leads to anomalous moistening of the mid- and upper-troposphere, centered at around $400~$hPa, that is then transported towards SZ via the TEJ.
The TEJ is also enhanced further southwest of NI such that the propagation over the Arabian Sea and the Sahel Zone is guided by an anomalously strong flow of easterly winds (Fig.~\ref{fig:wind_background}~d,j).
We applied a Lagrangian trajectory analysis \citep{Sprenger2015} to investigate the number of days it takes to transport the moist anomalies from NI to SZ (see SI Sec.~\ref{sec:lagr_trajs_SI}, Fig.~\ref{fig:lagr_trajs}). The analysis confirmed that the moist anomaly takes around 6--9 days to arrive over SZ, consistent with the propagation pathways in Fig.~\ref{fig:propagation_olr}~a-e.

For timepoints around day 0, we observe a weak orographic forcing over the Ethiopian Highlands and the western coast (Fig.~\ref{fig:SZ_rainfall}). We also see an enhancement of the TEJ over the Sahel (Fig.~\ref{fig:SZ_convection}~a) but there is no substantial increase in moisture at around $400~$hPa (Fig.~\ref{fig:SZ_rainfall}~a,b).
These patterns change around 1--2 days after the arrival of the anomalous moisture in the upper troposphere.
We now observe anomalous heavy rainfall over the Sahel (Fig.~\ref{fig:SZ_rainfall}~c,d). The rainfall extends over the full zonal and meridional range of the Sahel from  $-5\degree{E}-40\degree{E}$; $5\degree{N}-15\degree{N}$ with two peaks in intensity, one is over the Ethiopian Highlands, one over the Northwest African coast (Fig.~\ref{fig:SZ_rainfall}~c).
Composites of Strong Propagation MSDs show only marginally enhanced intensity at the region of the upper-level TEJ at $200\,$hPa (compare the zonal circulation at day 0, Fig.~\ref{fig:SZ_rainfall}~a, with day +12, Fig.~\ref{fig:SZ_rainfall}~c).
The upper-level westerlies do not vary substantially with respect to the climatology, consistent with \citep{Lemburg2019}.
Instead, anomalous moistening happens from near the surface up to about $400~$hPa (Fig.~\ref{fig:SZ_rainfall}~c,d).
Hence, it is not the local variability of the TEJ on weekly to subseasonal timescales that modulates the Sahel rainfall; rather its variability on interannual timescales. The strength of the teleconnection also, therefore, does not depend on the local strength but the positioning of the TEJ.

Already at day 0, we find weak anomalous orographic forcing over the Ethiopian Highlands but further moisture incursion is restricted by the Simien Mountains (around $40\degree{E}$ in Fig.~\ref{fig:SZ_rainfall}~a).
However, as we see on day +12, the upper-tropospheric moisture anomalies can readily pass over this mountain chain barrier, carried by the TEJ (Fig.~\ref{fig:SZ_rainfall}~c).
The anomalous moisture in the east ($>45\degree{E}$) is transported towards the west and the north of the Sahel Zone (Fig.~\ref{fig:SZ_rainfall}~c,d).
At this stage the center of the composite deep convection is around $20\degree{E}$ and $15\degree{N}$ (Fig.~\ref{fig:propagation_olr}~e, SI Fig.~\ref{fig:SZ_convection}).

The meridional cross-section on day 12 (Fig.~\ref{fig:SZ_rainfall}~d) shows an overturning circulation, where convection occurs slightly off the equator at around $10 \degree{N}$ and is extremely deep, extending to the tropopause.
This circulation overturns and leads to subsidence at both $5 \degree{N}$ and $15\degree{N}$.
The anomalous moisture brought to the mid- and upper-troposphere above the SZ by the TEJ leads to deep instability in a region already characterized by widespread convection during this time of year.
Further, the anomalous upper-level moisture supports sustained deep convection by reducing dry entrainment, ultimately leading to heavier widespread precipitation.
There is also a positive feedback, with the overturning circulation facilitating convergence in the lower levels, leading to the observed increase in near-surface humidity.
In particular, we observe enhanced integrated vapor transport (IVT) influx driven by the increased convective activity (Fig.~\ref{fig:ivf_flow}), which brings additional moisture from the Atlantic Ocean into the continent at levels around $900~$hPa (Fig.~\ref{fig:SZ_rainfall}~c, Fig.~\ref{fig:z900}).

\section{Discussion}
We have identified and explained a teleconnection pattern linking extreme rainfall events (EREs) on a continental scale between North India as part of the ISM and the Sahel Zone as part of the WAM.
We first demonstrated that the synchronization pattern of EREs between North India and the Sahel Zone is a robust feature of the boreal summer monsoon system.
The synchronization pattern is not a statistical artifact but is rather driven by anomalous large-scale atmospheric circulation, supported by the background state of the climate system.
Our work provides a physical explanation for the teleconnection between the Asian and the African Monsoon domains.

We find that the synchronization is initiated by anomalously strong convection in the northwest of India, which is associated with an enhanced active phase of the BSISO.
During La Ni\~na conditions there is typically a stronger South Asian monsoon (via the Walker circulation) and, driven by the BSISO, it is more likely for EREs to arise in northwest India \citep{Strnad2023}.
Importantly, these EREs substantially moisten the upper troposphere over NI via detrainment associated with deep convection.
Subsequently, the anomalous upper-level moisture is advected over the Arabian Sea and Sahel by the TEJ -- itself strengthened by La Ni\~na conditions. Moist air aloft over the Sahel provides conditions that are then favorable for supporting and sustaining deep convection (and hence EREs) there.
Hence, the strength of this teleconnection depends on the positioning of the TEJ and not on its regional intensity over SZ.

While prior investigations have provided detailed insights into intraseasonal precipitation variability at a regional level in North India \citep{Malik2010, Stolbova2014, Boers2019, Hunt2021, Hunt2022} and parts of the Sahel \citep{Gleixner2017, Lemburg2019, Vashisht2021}, our study integrates and consolidates these findings into a broader context. This synthesis enhances our comprehension of the underlying physical mechanisms, thereby holding the potential to enhance seasonal and sub-seasonal forecasts using windows of opportunities \citep{Mariotti2020} during boreal summer in the tropical monsoon domain.

Still, some open questions remain. Firstly, the propagation pathway shown in this study was uncovered by using OLR as a proxy for intense rainfall. OLR has the advantage of being directly measurable and therefore a reliable variable.
However, it is not a direct measure of precipitation.
Secondly, the local causes of the enhanced rainfall in the Sahel Zone remain unclear. Deep convective precipitation over the SZ relies on the interaction of several complex processes, including interactions of mesoscale convective systems with the East African Jet at around $600~$hPa and the TEJ.
Thirdly, the role of the CGT in the synchronization between North India and the Sahel Zone is not fully understood. We find some patterns indicating a partial modulation by the CGT, but its effect seems to be minor.
Further, future research could better investigate the Weak Propagation propagation cluster. As the previous analysis suggests that the Strong Propagation is the dominant process, the Weak Propagation could be a combination of multiple and presumably more local interactions.

There are multiple ways to extend this work.
One first step would be to test the occurrence of the synchronization pattern in current general circulation models (GCMs). On the one hand, an analysis of the teleconnection pattern in GCMs could provide insights into the robustness of the synchronization pattern. On the other hand, as precipitation dynamics and teleconnections are not always well reproduced in the current GCMs \citep{Boyle2010, Hess2022, IPCC_2023}, the comparison could help to identify potential biases in the models.
Another next step could be to integrate the knowledge about the uncovered teleconnection in current operational forecast systems. Of particular interest will be the representation of these synchronizations in deep learning global forecasts \citep{Lam2023, Bi2023, Lessig2023}, which are better capable of capturing complex non-linear relationships in the data.

\section*{Data Availability}
Precipitation data was taken from the MSWEP dataset (https://www.gloh2o.org/mswep)\cite{Beck2019}.
Datasets for the composite analysis from 1979 till date were taken from Copernicus Climate Change Service (C3S) (https://cds.climate.copernicus.eu/cdsapp\#!/dataset/reanalysis-era5-pressure-levels?tab=overview) \cite{ERA5}.
Plots were generated using the Cartopy library \cite{Cartopy}.
\section*{Code Availability}
The code for generating and analyzing the networks is made publicly available under \cite{CodeClimnet}. The analysis uses the geoutils package \citep{geoutils}.
The code for reproducing the analysis of the network communities and the data analysis in this paper is publicly available under \cite{CodeMonsoonSync}.
Plots were generated using the Cartopy \citep{Cartopy} and the Metpy v1.6 libraries \citep{metpy}.

\section*{Acknowledgements}
F.S. and B.G. acknowledge funding by the Deutsche Forschungsgemeinschaft (DFG, German Research Foundation) under Germany's Excellence Strategy - EXC number 2064/1 - Project number 390727645. F.S. thanks the International Max Planck Research School for Intelligent Systems (IMPRS-IS) for supporting his PhD program. N.B. acknowledges funding by the Volkswagen Foundation, the European Union's Horizon 2020 research and innovation program under the Marie Sklodowska-Curie grant agreement No. 956170, as well as from the European Union's Horizon Europe research and innovation program under grant agreement No. 101137601. K.M.R.H. is supported by a NERC Independent Research Fellowship (MITRE; NE/W007924/1).

\section*{Author's contribution}
F.S. and B.G. conceived and designed the study. F.S. conducted the analysis. F.S. and B.G. prepared the manuscript. F.S., K.M.R.H., N.B., and B.G. discussed the results and edited the manuscript.

\section*{Competing Interests}
The authors declare that they have no competing interests.
\clearpage
\newpage
\bibliography{./library.bib}

\newpage
\appendix
\renewcommand{\thefigure}{S\arabic{figure}}
\renewcommand\thesection{SI~\arabic{section}}
\setcounter{figure}{0}
\section*{Supplementary Information}

In this Supplementary Information, we provide additional information on the data and methods used in this study. We also provide additional figures that support the main text. The Supplementary Information is organized as follows:
In Sec.~\ref{sec:climate_network_si}, we provide additional information on the climate network used in this study.
In Sec.~\ref{sec:community_detection_si}, we provide additional information on the basic pattern of the synchronization between North India and the Sahel Zone.
In Sec.~\ref{sec:yearly_fluctuations_si} we provide additional information on the yearly fluctuations of the synchronization between North India and the Sahel.
In Sec.~\ref{sec:grid_search_mvpca} we outline the grid search for the optimal parameters of the multivariate PCA.
In Sec.~\ref{sec:TEJ_SI}, we give additional information on the Tropical Easterly Jet (TEJ) and its characteristics.
In Sec.~\ref{sec:CGT_SI}, we outline additional information on the Circumglobal Teleconnection (CGT) and its characteristics.
\ref{sec:PDO_SI} provides additional information on the Pacific Decadal Oscillation (PDO) and its characteristics.
The characteristics of the Somali Jet are shown in Sec.~\ref{sec:somali_jet_SI}.
We show the characteristics of the convective situation in the Sahel during day +12 in Sec.~\ref{sec:convection_SZ_SI}.
In Sec.~\ref{sec:bsiso_phases_SI}, we provide additional information on the Boreal Summer Intraseasonal Oscillation (BSISO) and its characteristics.
A quasi-geostrophic (QG) analysis of the days of the synchronization is shown in Sec.~\ref{sec:qg_analysis_SI}.
In section \ref{sec:lagr_trajs_SI}, we provide additional information on the propagation pathways of the synchronization between North India and the Sahel Zone as they are estimated by a Lagrangian trajectory model.

\section{Climate Network of Synchronous Rainfall Events} \label{sec:climate_network_si}
\paragraph{Event Synchronization}
\begin{figure}[!htb]
    \centering
    \includegraphics[width=.6\textwidth]{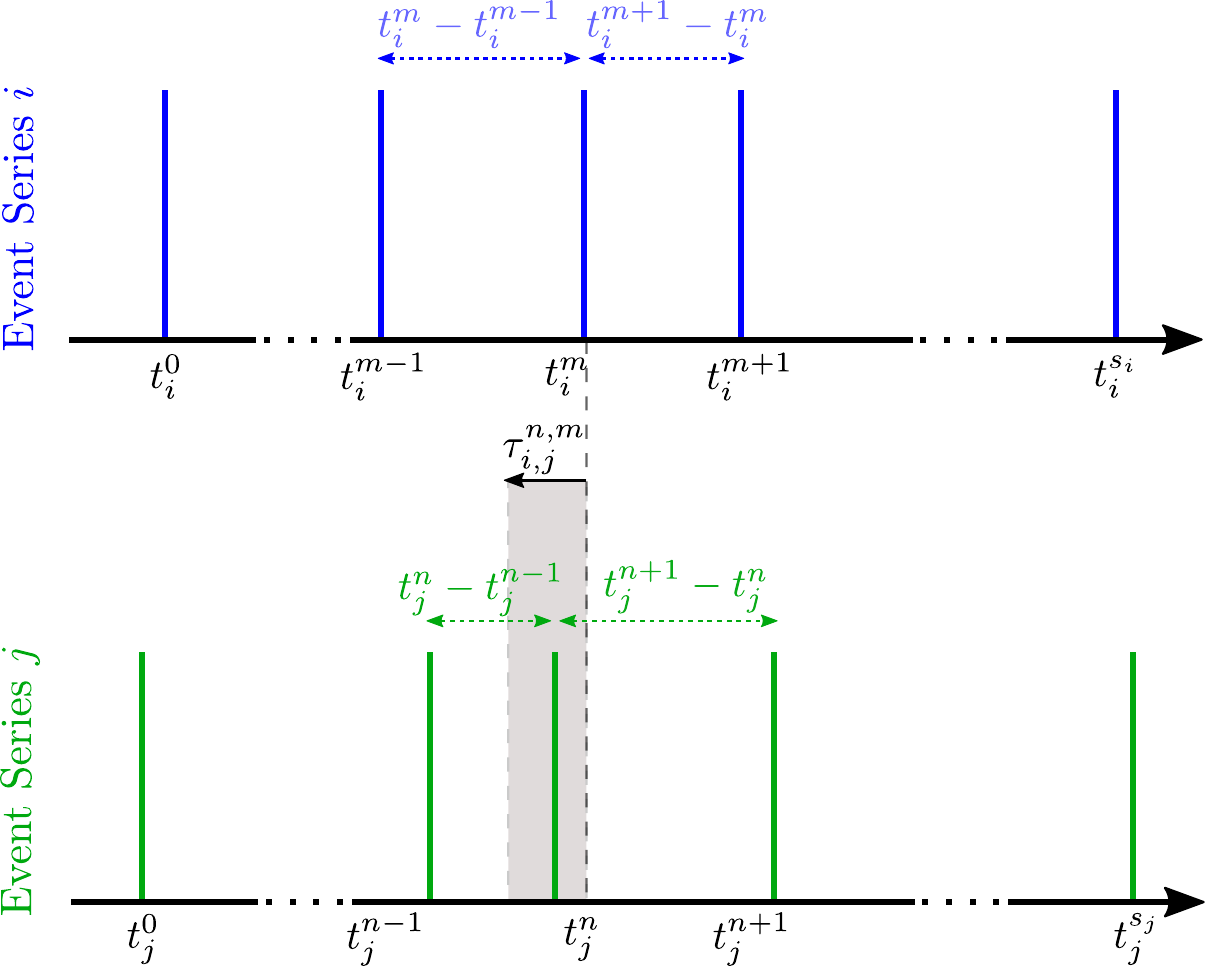}
    \caption{\textbf{Scheme of time-delayed Event Synchronization.}
        The blue (green) bars denote events that happen at the locations in $i$ ($j$) at the time points $t_i^m$ ($t_j^n$). $m$ ($n$) denotes the event number at location $i$ ($j$).
        The point in time  $ t_i^m $ is identified as synchronous to $ t_j^n $ since the time difference between $ t_i^m $ and $ t_j^n $ is within the allowed range given by $\tau_{ij}^{m,n}$ according to the definition expressed in equation \ref{eq:tau_mn}. Not shown here is the maximum delay $\tau_{\text{max}} $. }
    \label{fig:event_sync}
\end{figure}
To quantify the level of synchronization between pairs of time series, we employ the Event Synchronization algorithm \cite{Quiroga2002}. This algorithm counts the number of events that occur at the same time between pairs of event sequences ${e^m_i}{m=1}^{s_i}$ and ${e^n_j}{n=1}^{s_j}$, where $s_i$ ($s_j$) represents the total number of events at location $i$ ($j$), and $e^m_i$ ($e^n_j$) denotes the timing of an event in location $i$ ($j$). The delay between an event $e^m_i$ in location $i$ and an event $e^n_j$ in location $j$ is denoted as $d_{ij}^{m,n} = e_i^m - e_j^n$. The set $D_{ij}(e^m_i, e^n_j)$ is defined as the set that contains the four neighboring events of $e^m_i$ and $e^n_j$:

\begin{equation}
    D_{i,j}^{m,n} = \left\{ d_{i,i}^{m,m-1}, d_{i,i}^{m,m+1},
    d_{j,j}^{n,n-1}, d_{j,j}^{n,n+1},
    2 {\tau_{\text{max}}} \right\},
\end{equation}

The dynamical delay, denoted as $\tau_{i,j}^{m,n}$, is calculated as half of the minimum time difference between subsequent events in both time series around event $e_i^m$ and $e_j^n$. It captures a small time delay between events and is limited to a maximum value of $\tau_{\text{max}}$:

\begin{equation}
    \tau_{ij}^{m,n} = \frac{1}{2} \min_{\forall d \in D_{ij}^{m,n}} d \; .
    \label{eq:tau_mn}
\end{equation}

The parameter $\tau_{\text{max}}$ represents the maximum allowable time delay between two events, capturing both short and long-spatial range interactions. In this study, we set $\tau_{\text{max}}$ to 10 days. The event synchronization strength $R_{i,j}$ between locations $i$ and $j$ is calculated as the sum of synchronous time points between all pairs of event sequences ${e^m_i}{m=1}^{s_i}$ and ${e^n_j}{n=1}^{s_j}$:

\begin{align}
    R_{i,j} & = \sum_{m=1}^{s_i} \sum_{n=1}^{s_j} S_{i,j}^{m,n} \hspace{1cm} \text{where} \hspace{0.5cm}
    S_{i,j}^{m,n} = \begin{cases}
                        1 & \hspace{.5cm} 0< d_{ij}^{m,n} < \tau_{i,j}^{m,n} \;, \\
                        0 & \hspace{.5cm}  \text{otherwise}\;.
                    \end{cases}
\end{align}

We count blocks of consecutive events as a single event, positioned at the time of the first event. This avoids a dynamical delay value of $1/2$, which would lead to a case where two sequentially occurring events are not considered synchronous. The complete event synchronization scheme is illustrated in Fig.~\ref{fig:event_sync}.

\paragraph{Constructing the network.}
\begin{figure}[!htb]
    \centering
    \includegraphics[width=1\linewidth]{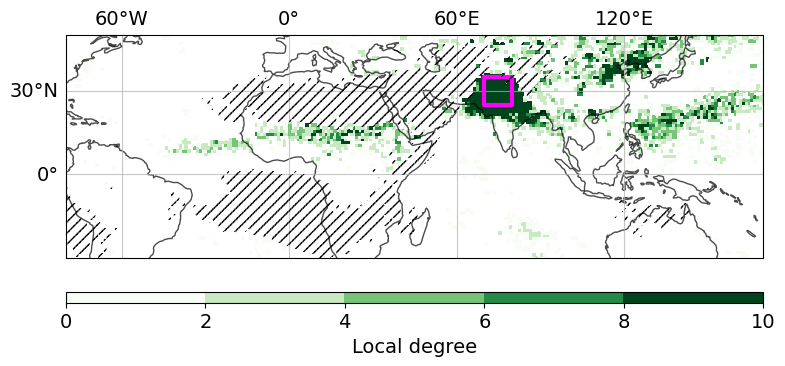}
    \caption{\textbf{Event Synchronization based network links from North India box.} We investigate the occurrence of rainfall extremes in the core monsoon season from June --- September (JJAS). The network links the occurrence of Extreme Rainfall Events  (EREs) that occur in the box in North India (NI) [70-85°E; 28-33°N] (pink rectangle) to synchronously occurring locations using an event-synchronization \citep{Quiroga2002} based climate network \citep{Dijkstra2019} approach (compare \citep{Boers2019, Strnad2023}) using a dynamical time lag of $\tau_{max}=10~$days.
        Green dots denote locations in which ERES occur statistically significantly synchronously or within the time lag $\tau_{max}$ after the occurrence of EREs within the North India box (pink rectangle).
        The color bar denotes the number of connections from a certain location to a location in NI.
        The hatched lines indicate regions with only very little rainfall in JJAS and these locations are excluded from the analysis.
    }
    \label{fig:climate_network}
\end{figure}
The adjacency matrix $\mathbf{A}$ of a network represents the connections between nodes and defines the network's underlying topology. It is an $N\times N$ matrix, where $A_{i,j}=1$ indicates that events at location $i$ are statistically significantly followed by events at location $j$. To determine the statistical significance, we perform a null-model test. The null hypothesis assumes that the observed $R_{i,j}$ value arises from a pair of randomly generated event sequences $e_i', e_j'$ with the same number of events $s_i, s_j$ as the observed sequences. We construct surrogate event sequences by randomly and uniformly distributing events. We consider $e_i$ to be significantly synchronous with $e_j$ if their corresponding $R_{i,j}$ value exceeds the 95th percentile of $R_{i', j'}$ values obtained from 2000 pairs of surrogate event sequences $e_i', e_j'$. When a significant $R_{i,j}$ value is found, we establish an edge from node $n_i$ to $n_j$ and set $A_{i,j}=1$.

An example of network links of locations in North India between $70\degree{E}$ and $80\degree{E}$ and $25\degree{N}-35\degree{N}$ is shown in Fig.~\ref{fig:climate_network}. For this region, we observe a synchronization of rainfall extremes in North India with the North-East China region around the Yellow River basin, the Western Pacific above the Philippines, and the Sahel Zone.
The connection towards the Yellow River basin has been analyzed previously \citep{Gupta2022}, in this work the focus will be set on the synchronization between the monsoon system in North India and the Sahel Zone.

\section{Community Detection in the tropics and subtropics} \label{sec:community_detection_si}
\begin{figure}[!htb]
    \centering
    \includegraphics[width=1.\linewidth]{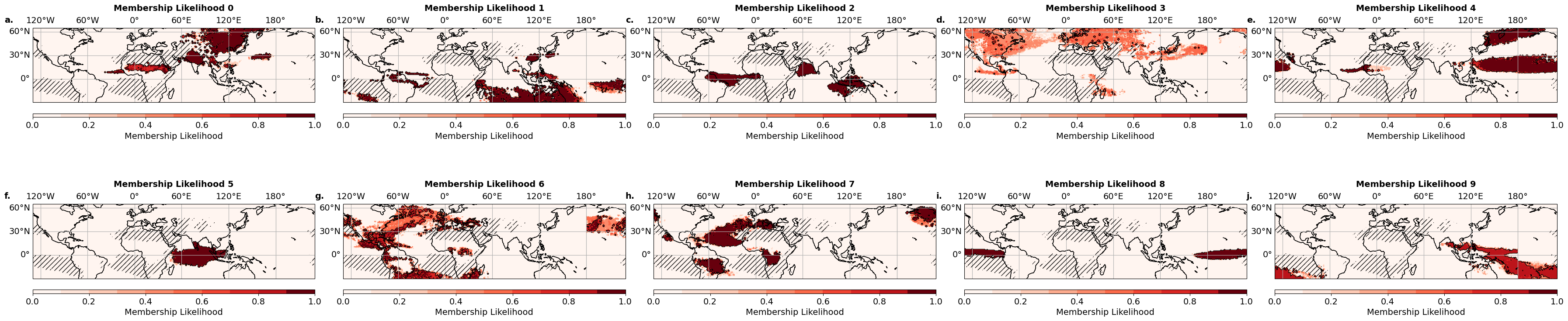}
    \caption{\textbf{Membership Likelihoods for different communities.} Using the heuristic outlined in sec. \ref{fig:basic_pattern} we find $10$ communities.
        The color bar shows the membership likelihood of a respective community. $100$ independent runs of the community detection algorithm have been used for this analysis. We find that some communities show a quite stable occurrence, some others are more fluctuating. However, for this work, we are mainly interested in the North India - Sahel zone community which is one of the most stable communities of the algorithm.
    }
    \label{fig:msl_all}
\end{figure}

\section{Yearly Fluctuations of the Synchronization} \label{sec:yearly_fluctuations_si}
\begin{figure}[!htb]
    \centering
    \includegraphics[width=.8\linewidth]{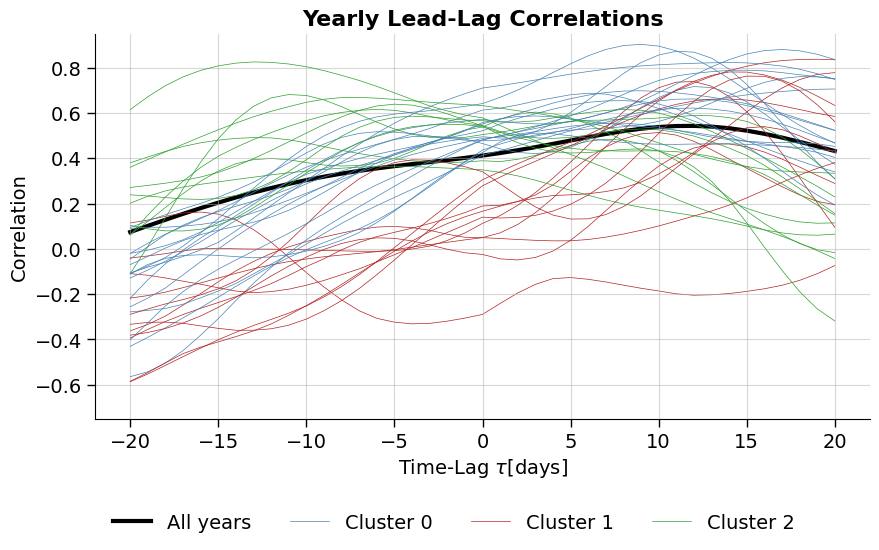}
    \caption{\textbf{Yearly fluctuations of the synchronization.}
        The yearly lead-lag correlations of extreme rainfall events (EREs) between North India (NI) and the Sahel Zone (SZ) are displayed. The lead-lag correlation is computed using the counts of EREs per year in NI (SZ). The individual yearly lead-lag correlations are clustered using a Gaussian Mixture model. This indicates that the lead-lag correlation is not constant but varies between different years.}
    \label{fig:yearly_lead_lag}
\end{figure}

This section investigates the yearly fluctuations of the synchronization between North India and the Sahel Zone. The different years are clustered using a Gaussian Mixture model. This indicates that the lead-lag correlation is not constant but varies between different years. We find that the synchronization is not constant but varies between different years (Fig.~\ref{fig:yearly_lead_lag})

\section{Grid search for optimal parameters of the Multivariate PCA} \label{sec:grid_search_mvpca}
\begin{figure}[!htb]
    \centering
    \includegraphics[width=.6\linewidth]{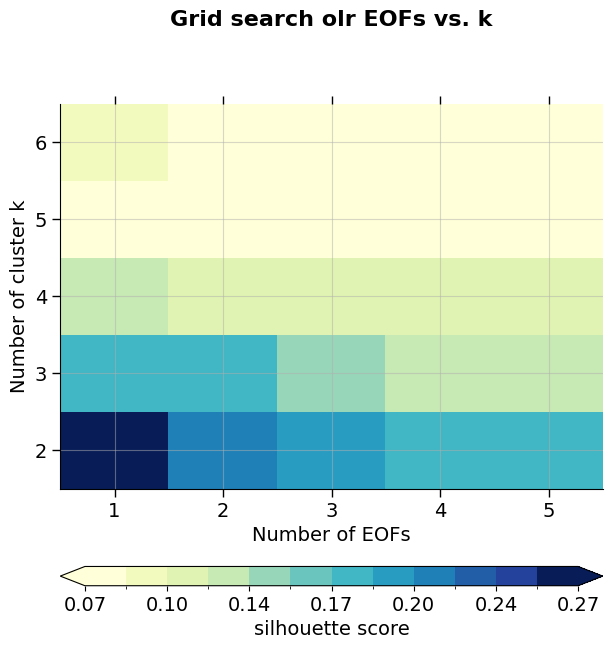}
    \caption{\textbf{Grid search for optimal parameters of the Multivariate PCA.}
        The optimal parameters for the Multivariate PCA are found by a grid search. The grid search is performed by varying the number of EOFs and the number of clusters. The optimal parameters are found by maximizing the Silhouette Score.}
    \label{fig:grid_search_mvpca}
\end{figure}

The optimal parameters for the Multivariate PCA are found by a grid search. The grid search is performed by varying the number of EOFs and the number of clusters. The optimal parameters are found by maximizing the Silhouette Score (Fig.~\ref{fig:grid_search_mvpca}).

\section{Tropical Easterly Jet Index} \label{sec:TEJ_SI}
This section presents the characteristics of the Tropical Easterly Jet (TEJ) according to the definition by \cite{Huang2019}.
The TEJ is a prominent feature of the Asian summer monsoon circulation, driven by the meridional thermal contrast between the Asian landmass and the Indian Ocean. It is further intensified by the elevated heating over the Tibetan plateau. The variability of the TEJ substantially influences climate patterns in the tropical and subtropical domain over Asia and the Sahel zone and its surrounding regions.

\begin{figure}[!htb]
    \centering
    \includegraphics[width=\linewidth]{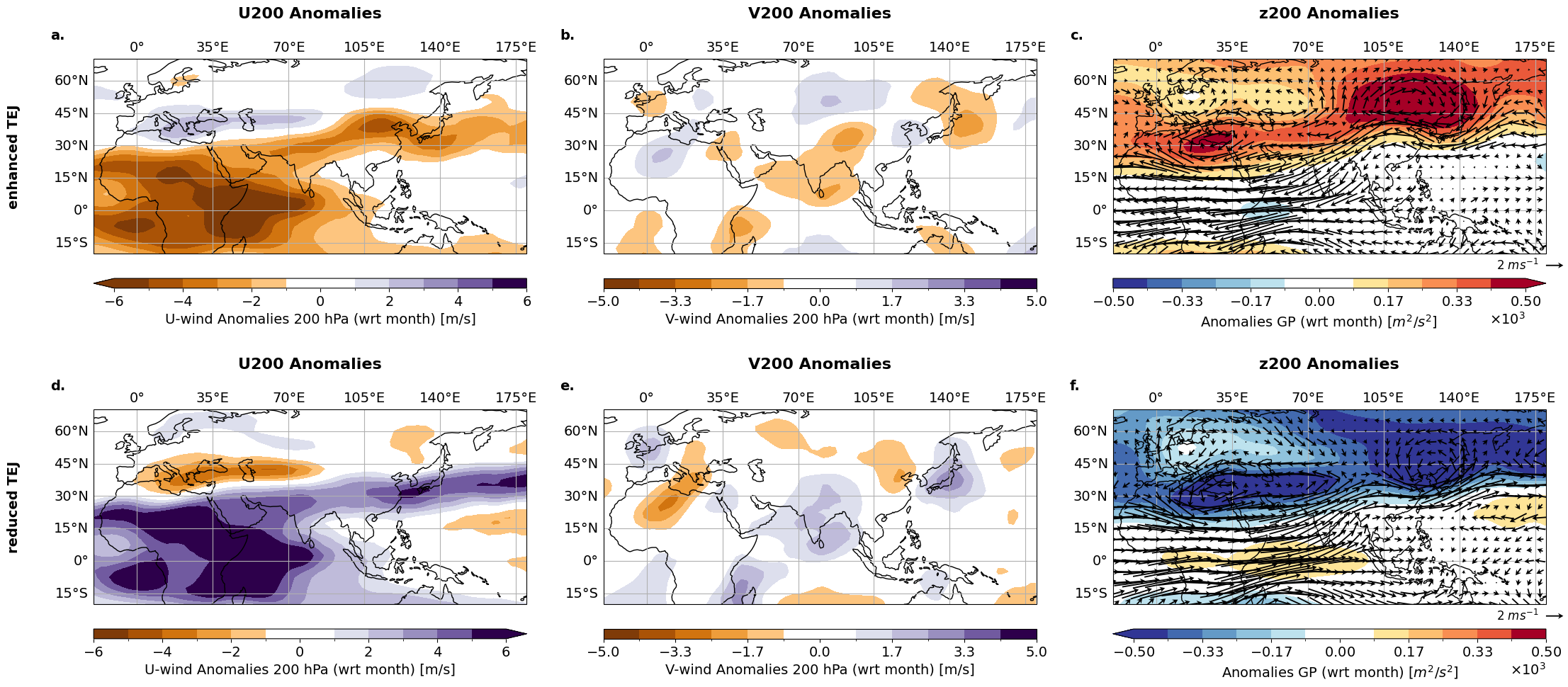}
    \caption{\textbf{Illustration of enhanced/reduced TEJ phases.}
        Composites of different phases of the TEJ are shown. The first row shows the enhanced state, and the second row the reduced state.
        The first column (\textbf{a}, \textbf{d}) displays the U200 anomalies (with respect to month), the second column (\textbf{b}, \textbf{e}) showcases the V200 anomalies, and third row (\textbf{c}, \textbf{f}) exhibits the Geopotential anomalies at $200$ hPa overlapped by the wind field anomalies at $200~$hPa, as identified by the TEJI metric described in Sec.~\ref{sec:data}. The magenta rectangle in pannel \textbf{a} visualizes the box that is used to define the TEJI.}
    \label{fig:def_tej_pattern}
\end{figure}

\paragraph{Pointwise spatial correlation analysis of wind fields} \label{sec:pointwise_correlation_si}
We uncover that multiple potential drivers come together when the probability of observing the delayed synchronization between North India and the Sahel Zone is notably increased (Fig.\ref{fig:likelihoods_bsiso_enso_tej}).
We find a substantially higher frequency of synchronizations during an active Boreal Summer Intraseasonal Oscillation (BSISO) in phases 5-8, conditioned on a  La Ni\~na background state and an intensified Tropical Easterly Jet (TEJ) in a westerly direction (Fig.~\ref{fig:likelihoods_bsiso_enso_tej}~c).
The pronounced role of the TEJ in establishing the synchronization is underlined by the fact that in its absence there are hardly any synchronizations observed regardless of the respective ENSO state and the BSISO phases (Fig.~\ref{fig:likelihoods_bsiso_enso_tej}~b,d,f). Similarly, El Ni\~no-like conditions substantially decrease the likelihood of observing synchronizations (Fig.~\ref{fig:likelihoods_bsiso_enso_tej}~a,b). Further, synchronizations during inactive BSISO are distributed around the null model (grey bars in Fig~\ref{fig:likelihoods_bsiso_enso_tej}) and are not statistically significant.
We estimate the relation between BSISO phases (as defined by \cite{Kikuchi2012}) and concurrent EREs (Fig.\ref{fig:msl_all}~a) utilizing a conditional dependence test constructed as the conditional probability of synchronous rainfall events contingent upon (i) phase, (ii) active (inactive) BSISO, (iii) state of the El Ni~no Southern Oscillation (ENSO), and (iv) state of the TEJ.
This statistical analysis suggests the interpretation that La Ni\~na-like conditions establish the foundational state in which the BSISO is predisposed to propagate poleward over Northern India, as reported by \citep{Strnad2023}. This convective cloud over Northern India (NI) is then prone to be advected westwards through the summer circulation over NI in July and August enforced by an enhanced TEJ.
These three factors are now analyzed in detail.

\begin{figure}[!tb]
    \centering
    \includegraphics[width=1.\textwidth]{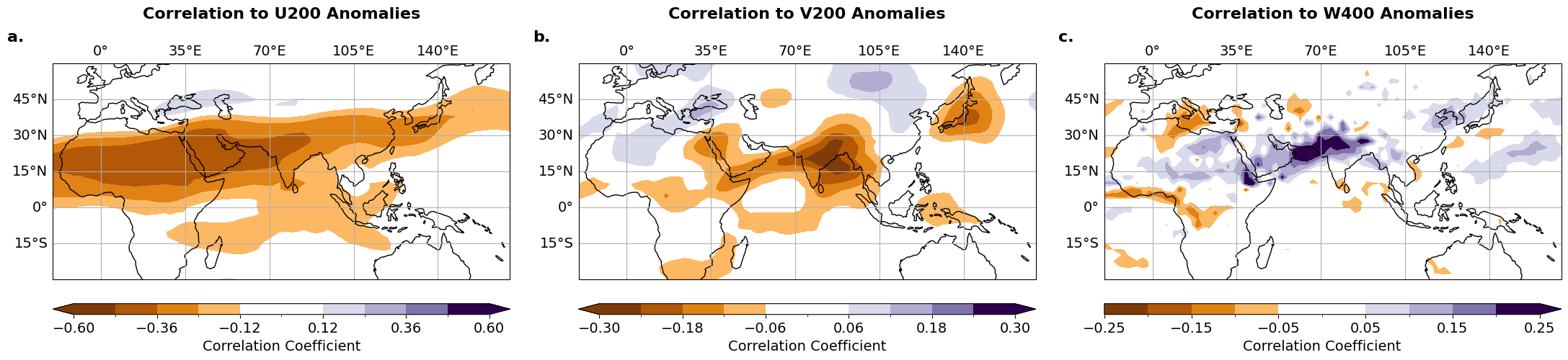}
    \caption{\textbf{Correlation of Synchronous Rainfall Index with wind fields.}
        \textbf{a} The pointwise Spearman correlation of the zonal u-wind anomalies at $200\,$hPa with the lagged synchronous rainfall index is shown.
        \textbf{b} Same as for \textbf{a} but for the meridional v-wind anomalies at $200~$hPa.
        \textbf{c} Same as for \textbf{a/b} but for the vertical velocity $\omega$ at $500~$hPa.
        Colored areas imply statistically significant correlations at the 99\,\% confidence interval using a two-sided Student's t-test.
    }
    \label{fig:tej_correlation}
\end{figure}

\section{Circumglobal Teleconnection (CGT)} \label{sec:CGT_SI}
\begin{figure}[!htb]
    \centering
    \includegraphics[width=\linewidth]{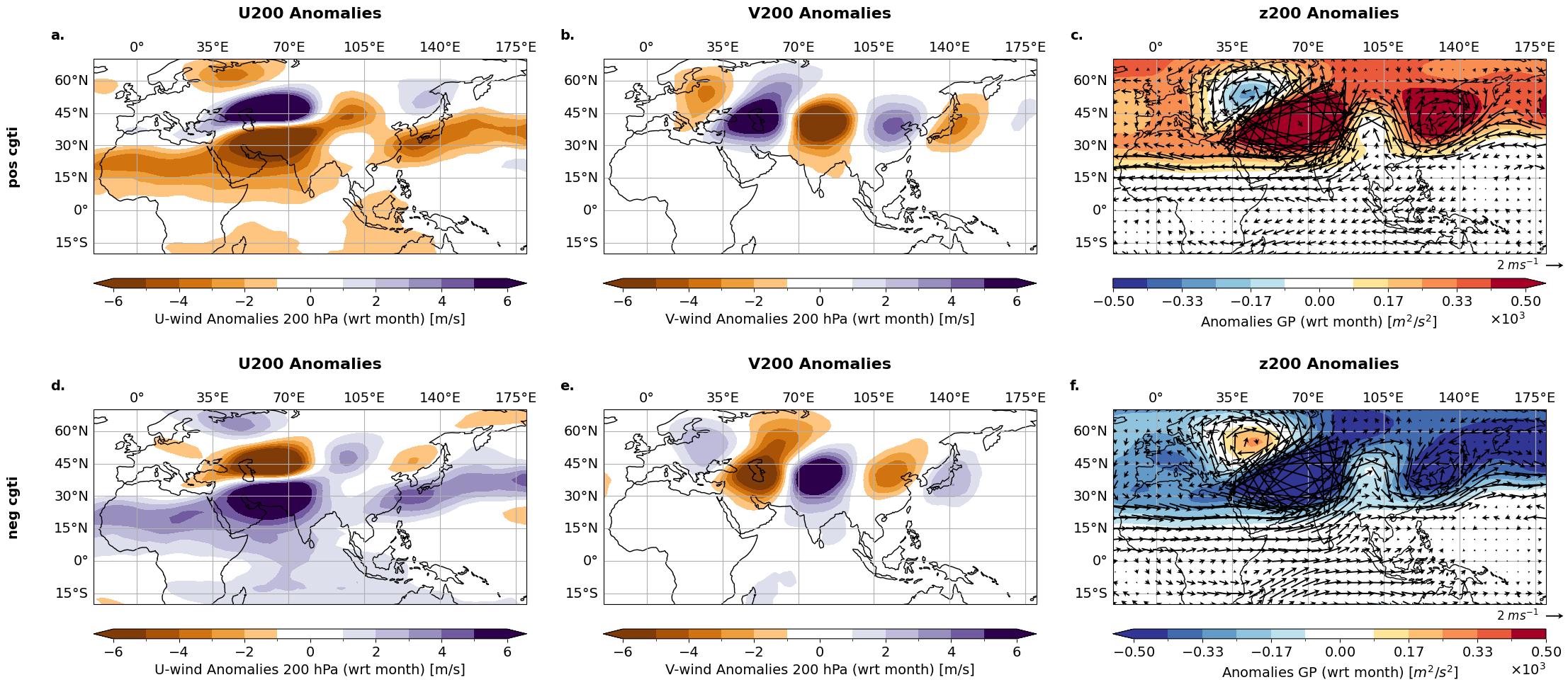}
    \caption{\textbf{Illustration of positive/negative CGT phases}.
        Composites of different phases of the CGT are shown. The first row shows the enhanced state, and the second row the reduced state.
        The first column (\textbf{a}, \textbf{d}) displays the U200 anomalies (with respect to month), the second column (\textbf{b}, \textbf{e}) showcases the V200 anomalies, and third row (\textbf{c}, \textbf{f}) exhibits the Geopotential anomalies at $200$ hPa overlapped by the wind field anomalies at $200~$hPa, as identified by the CGTI metric described in Sec.~\ref{sec:data}.
        The magenta rectangle in pannel \textbf{a} visualizes the box that is used to define the CGTI.}
    \label{fig:CGTI_phases}
\end{figure}

This section shows the characteristics of the Circumglobal Teleconnection Pattern (CGT) as defined by \cite{Ding2005}. The CGT is a wave pattern that is characterized by a zonal wavenumber of 5. It is a wave pattern that is known to be associated with the Asian Summer Monsoon and the North Atlantic Oscillation (NAO). The CGT is known to be associated with the Silk Road Pattern (SRP) \citep{Enomoto2003} and the East Asian Wave train pattern \citep{Ding2007}.
This characteristic wave pattern over Eurasia is shown in Fig.~\ref{fig:CGTI_phases} which resembles the wave pattern identified by \cite{Ding2007}.

\section{Connection to Pacific Decadal Oscillation (PDO)} \label{sec:PDO_SI}
\begin{figure}[!htb]
    \centering
    \includegraphics[width=\linewidth]{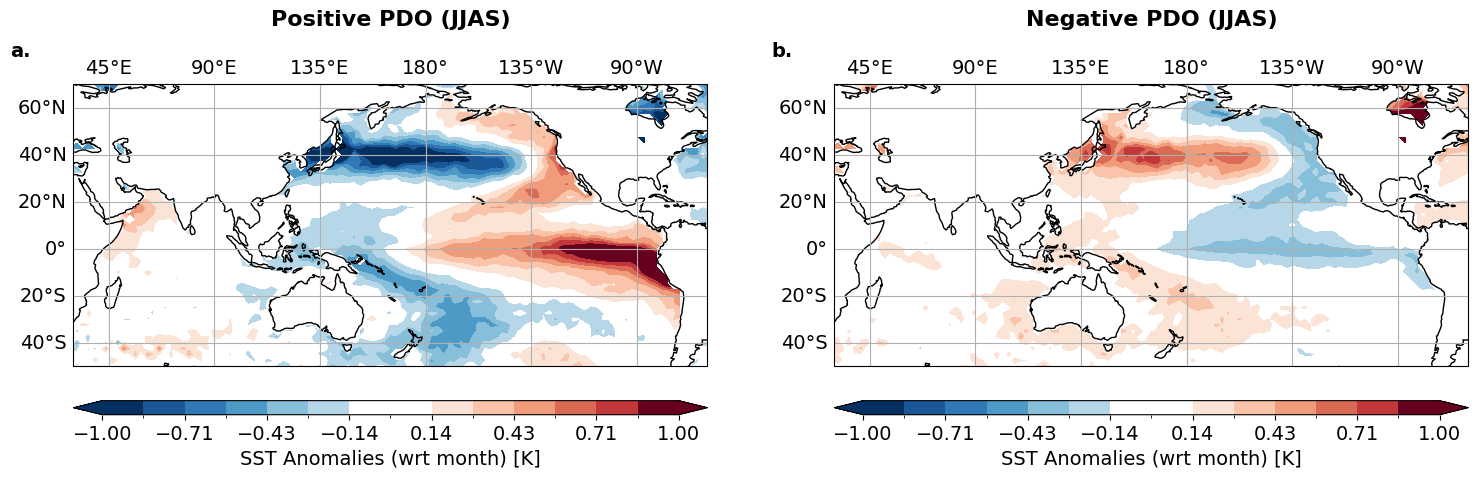}
    \caption{\textbf{Illustration of positive/negative PDO phases}.
        Composites of different phases of the PDO in JJAS are shown. Panel \textbf{a} shows the SST anomalies that are associated with the positive PDO phase, while panel \textbf{b} shows the SST anomalies that are associated with the negative PDO phase.}
    \label{fig:PDO_phases}
\end{figure}

We observe that the Pacific Decadal Oscillation (PDO) has a signature in the SST plots of the Strong Propagation (Fig.~\ref{fig:sst_background}~a). The PDO is a long-term climate pattern that is characterized by a positive and negative phase (Fig.~\ref{fig:PDO_phases}). The PDO is known to be connected with ENSO and influences the climate in the Indo-Pacific domain and the North American continent.
We observe that starting from around 1986, the phases of the PDO roughly align with the phases of the synchronization (Fig.~\ref{fig:yearly_occurrence_cluster_pdo})

\section{The Somali Jet} \label{sec:somali_jet_SI}
\begin{figure}[!htp]
    \centering
    \includegraphics[width=1.\linewidth]{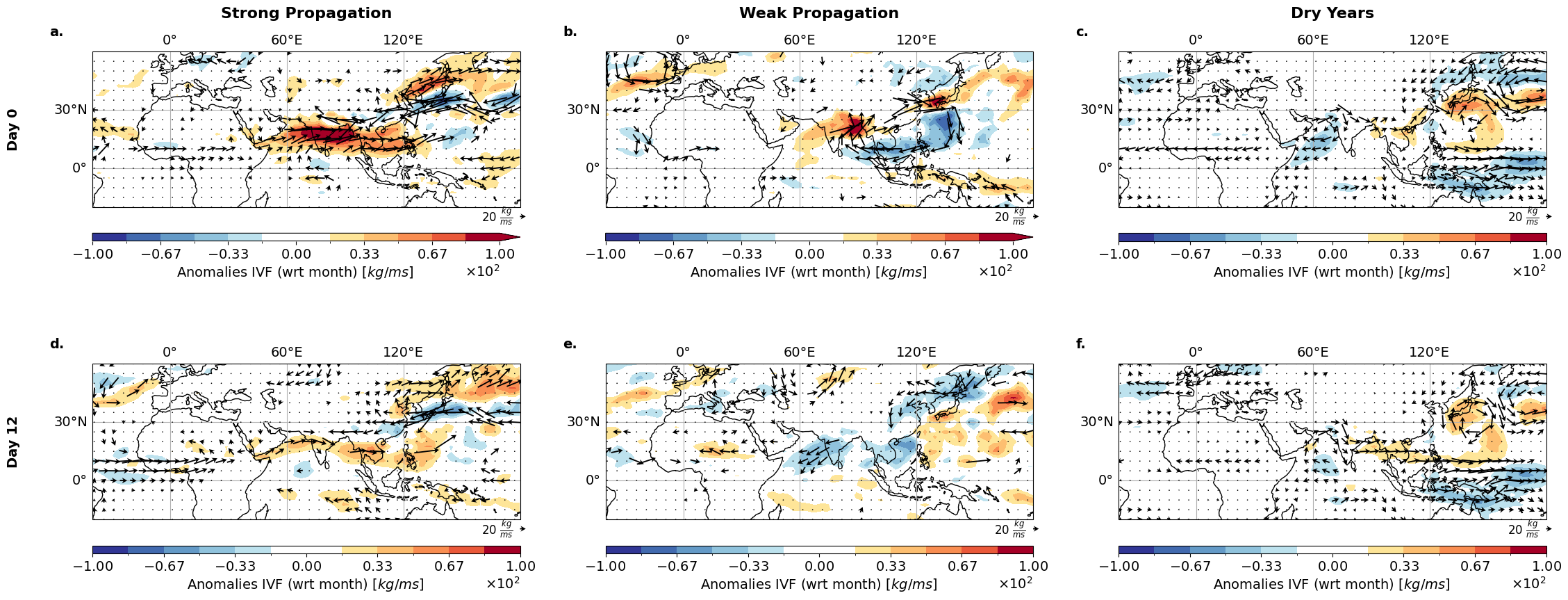}
    \caption{\textbf{Vertically integrated water vapor flux (IVT) for days of maximum synchronization for clustered regions.}
        Vertically integrated water vapor flux (IVF) in eastward and northward directions is visualized for the different clusters (\textbf{a,b}) and for the July-August background state for years with only a few synchronizations, called dry years.
        Only statistically significant anomalies at the 95~\% confidence level are shown.
        Day 0 denotes the days of maximum synchronization. The composite anomalies are computed with respect to the monthly climatology. Only IVT arrows that are significant at $95~\%$ level following the Student's t-test are plotted.
    }
    \label{fig:ivf_flow}
\end{figure}

The South Asian monsoon onset is marked by a rapid increase in rainfall over southwest India and is preceded by a reversal of winds in the lower troposphere. This occurs analogously to the formation of a low-level zonally oriented jet over the western Indian Ocean. The jet initially flows across the equator and then turns eastward near the East African coast into a zonal flow \citep{Findlater1969}. This jet is commonly known as the Somali jet. Its low-level southwesterly flow plays a crucial role in transporting moisture for the Indian and Asian Summer Monsoon. It precedes the onset of the monsoon over India and exhibits similar characteristics of rapid intensification and gradual retreat during the monsoon season \citep{Halpern2001, Masiwal2023}.
This flow can be visualized using Integrated Vapor Transport (IVT). We observe that for the MSDs the flow is enhanced over the Arabian Sea and the Bay of Bengal (Fig.~\ref{fig:ivf_flow}~a,b) bringing additional moisture towards the center and North of India. In dry years, consistently the flow is reduced (Fig.~\ref{fig:ivf_flow}~c).

\begin{figure}[!htp]
    \centering
    \includegraphics[width=1.\linewidth]{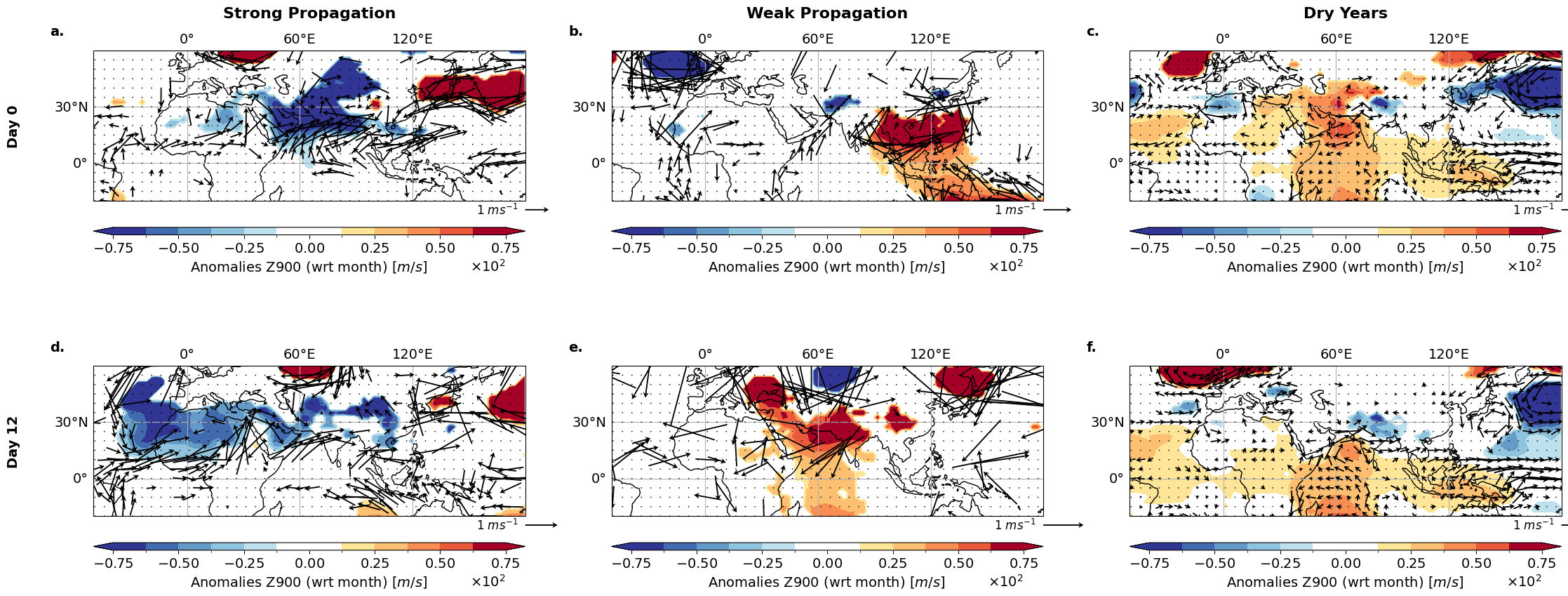}
    \caption{\textbf{Geopotential and horizontal wind fields at 900 hPa.}
        Wind fields and geopotential at $900$~hPa are visualized for the different clusters (\textbf{a,b}) and for the July-August background state for years with only a few synchronizations, called dry years.
        Only statistically significant anomalies at the 95~\% confidence level are shown.
        Day 0 denotes the days of maximum synchronization. The composite anomalies are computed with respect to the monthly climatology. Only wind arrows that are significant at $95~\%$ level following the Student's t-test are plotted.
    }
    \label{fig:z900}
\end{figure}

\section{Convection in SZ}\label{sec:convection_SZ_SI}
\begin{figure}[!tb]
    \centering
    \includegraphics[width=1.\linewidth]{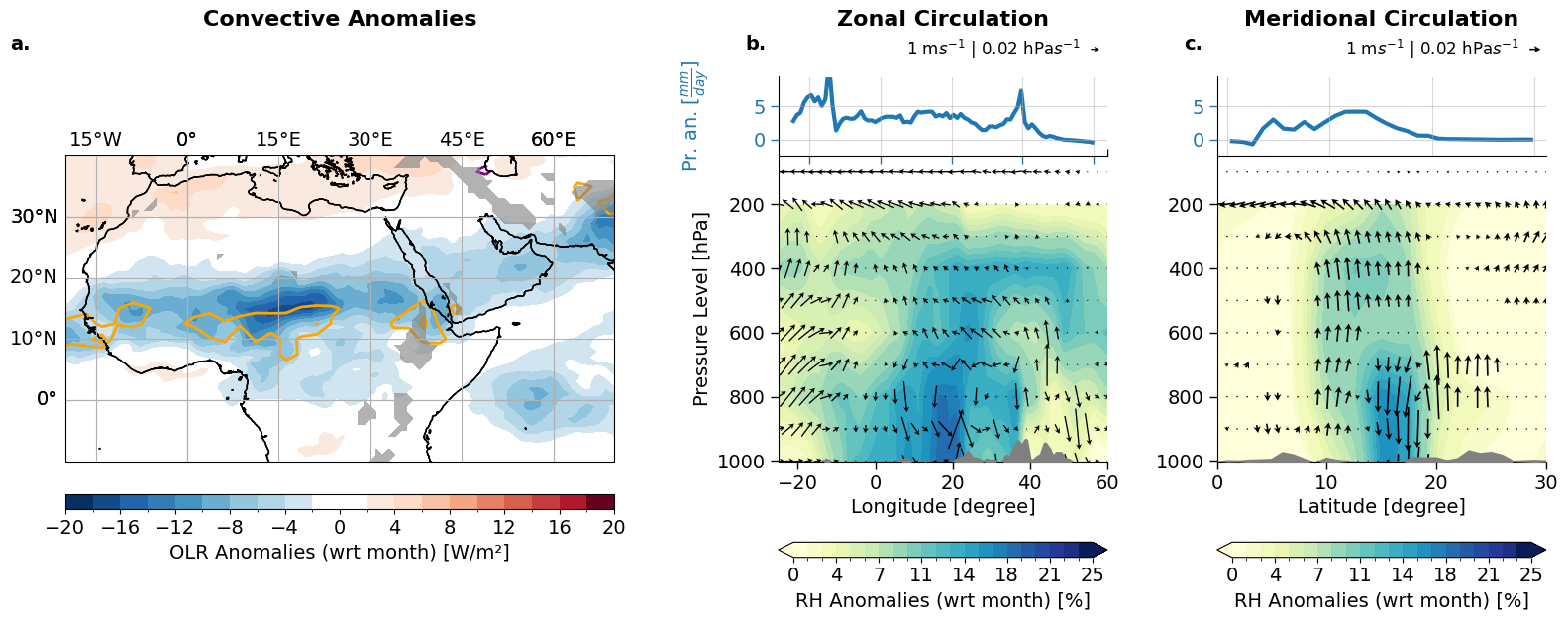}
    \caption{\textbf{Convective anomalies in North India.} Panel \textbf{a} shows the Outgoing Longwave Radiation (OLR) superimposed by the vertical velocities $\omega$ at $400~$hPa  (measured in Pa/s) by colored contours. Orange solid (purple dashed) contours denote anomalously rising conditions for regions with anomalies larger (smaller) than $3\,$Pa/s.
        Panel \textbf{b} visualizes the zonal circulation averaged between $20\degree{N}-30\degree{N}$ latitude, while panel \textbf{c} the meridional circulation, averaged between $75\degree{E}-85\degree{E}$ longitude.
        All composites are computed for the conditions on Day 12. The panel in \textbf{b} (\textbf{c}) is split into two parts: On top are displayed the meridionally (zonally) averaged precipitation anomalies while below the vertical circulation is shown by composites of relative humidity (RH) anomalies (shading) and wind fields (arrows). In \textbf{a} colored contours denote statistical significance at $95~\%$ confidence level using a two-sided t-test while grey contours visualize the orography.
        The wind fields in the zonal (meridional) circulation plots are estimated using the meridionally (zonally) averaged u (v) anomalies, measured in m/s, and the vertical velocity $\omega$ in the horizontal direction, measured in hPa/s. Only statistically significant arrows at $95\,\%$ confidence level using a two-sided t-test are shown.
    }
    \label{fig:SZ_convection}
\end{figure}

Analogously to the convection visualized in Fig~\ref{fig:NI_convection} for North India, we show the convection in the Sahel Zone (Fig.~\ref{fig:SZ_convection}) for the day +9 in the Sahel Zone for MSDs in the Strong Propagation cluster. The convection in the Sahel Zone is characterized by a strong rising motion in the atmosphere at around $400~$hPa. The centers of convection are around the Ethiopian Highlands at around $15\degree{E}, 12\degree{N}$ and the Western coast.

\section{Convection through BSISO} \label{sec:bsiso_phases_SI}
\begin{figure}[!tb]
    \centering
    \includegraphics[width=1.\linewidth]{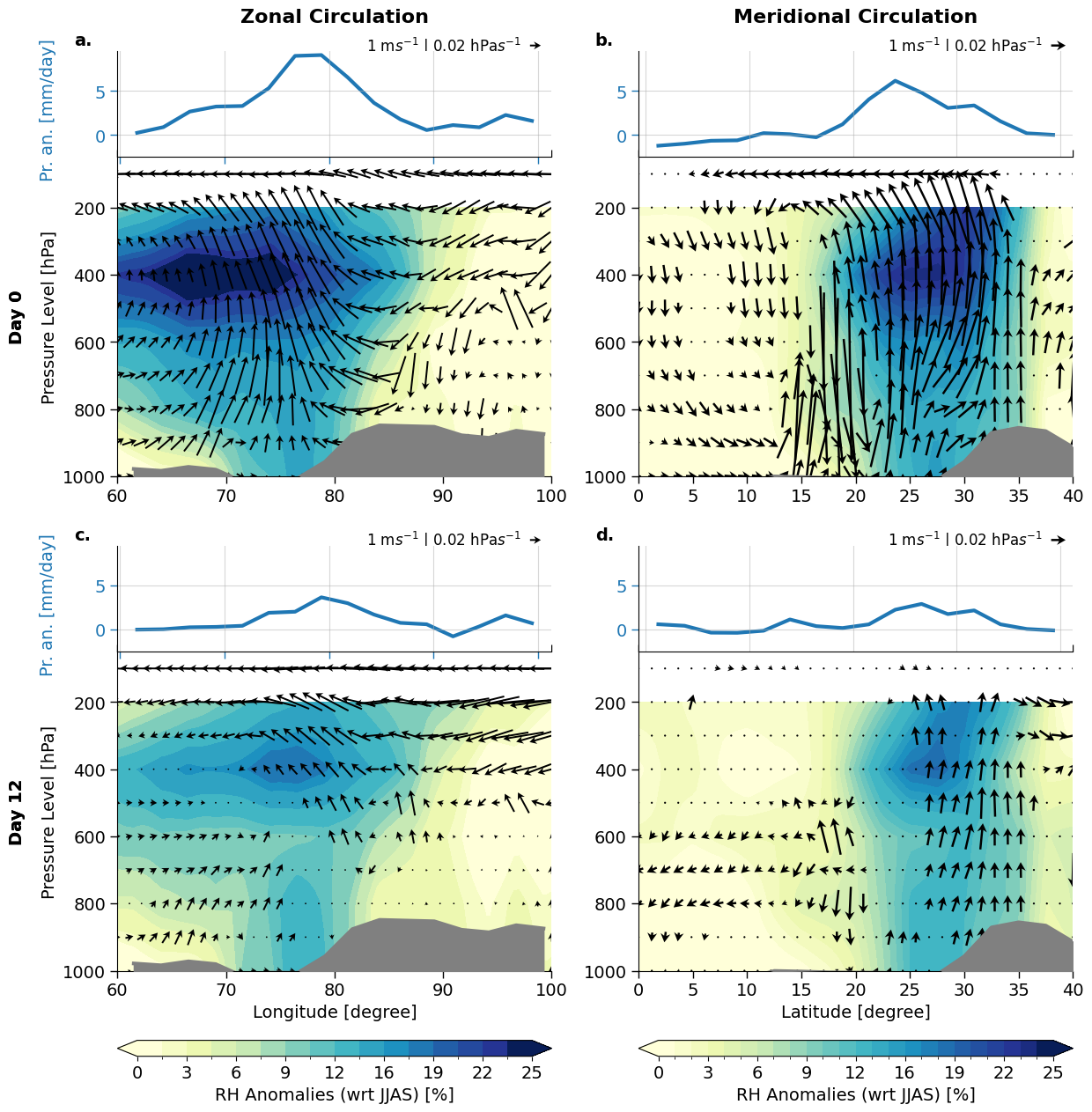}
    \caption{\textbf{Lagged precipitation occurrence in North India.} The first column visualizes the zonal circulation averaged between $20\degree{N}-30\degree{N}$ latitude, while the second column the meridional circulation, averaged between $75\degree{E}-85\degree{E}$ longitude. The first row shows the conditions on Day 0 and the second on Day 12, which is the day of the strongest correlation (see Fig.~\ref{fig:msl_all}~b). Each subplot consists of two panels. The top panel shows the meridionally (zonally) averaged precipitation anomalies (with respect to the month). The down panel shows the vertical circulation. Here, the colored contours show statistically significant composites of V-(U-) wind anomalies at the left (right) column. Grey contours visualize the orography.  The wind fields in the zonal (meridional) circulation plots are estimated using the meridionally (zonally) averaged u (v) anomalies, measured in m/s, and the vertical velocity $\omega$ in the horizontal direction, measured in hPa/s. Only statistically significant arrows at $95\,\%$ confidence level are shown.
    }
    \label{fig:bsiso_phases_convection}
\end{figure}

\begin{figure}[htbp]
    \centering
    \includegraphics[width=\linewidth]{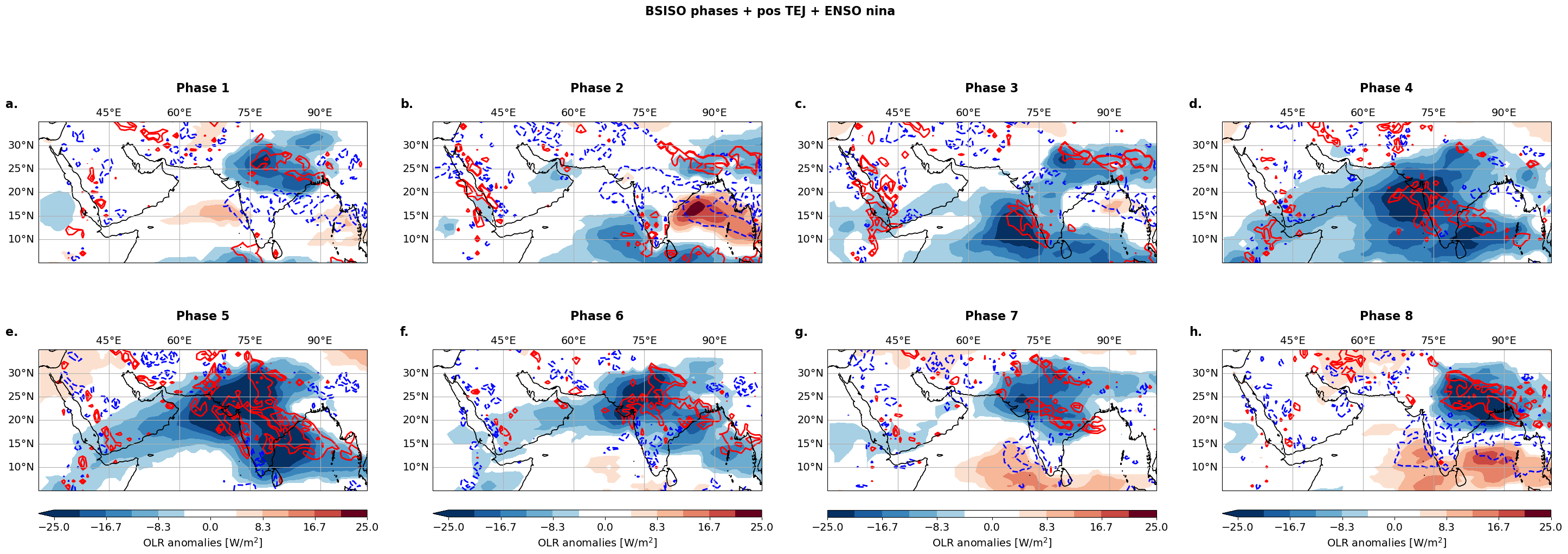}
    \caption{\textbf{Convection centers for BSISO phases.} For time points in La Ni\~na years with enhanced TEJ the phases 1-8 of the BSISO are shown in terms of OLR anomalies (color shading) and vertical velocity $\omega$ in contours where solid red (dashed blue) contours indicate anomalously upward (downward) motion.}
    \label{fig:bsiso_phases_LN_tej}
\end{figure}

The Boreal Summer Intraseasonal Oscillation (BSISO) is a prominent feature of the Asian Summer Monsoon. It is characterized by a northward propagation of convection over the Indian Ocean and the Maritime Continent. The BSISO  periodically occurs on time scales of around 30-60 days and is associated with the active and break events of the Asian Summer Monsoon \citep{Kikuchi2021}. The BSISO is characterized by eight phases, which are shown in Fig.~\ref{fig:bsiso_phases_convection}.
It has been shown that ENSO modulates substantially the propagation pathway of the BSISO \citep{Strnad2023}. In La Ni\~na years, the BSISO is more likely to propagate over the Indian subcontinent and the North Indian Ocean. Further, in La Ni\~na years the TEJ is enhanced. Therefore, also the convection centers and anomalies change and are more confined to the North and Northwest of India.
This is shown in Fig.~\ref{fig:bsiso_phases_LN_tej}.

\begin{figure}[!htb]
    \centering
    \includegraphics[width=\linewidth]{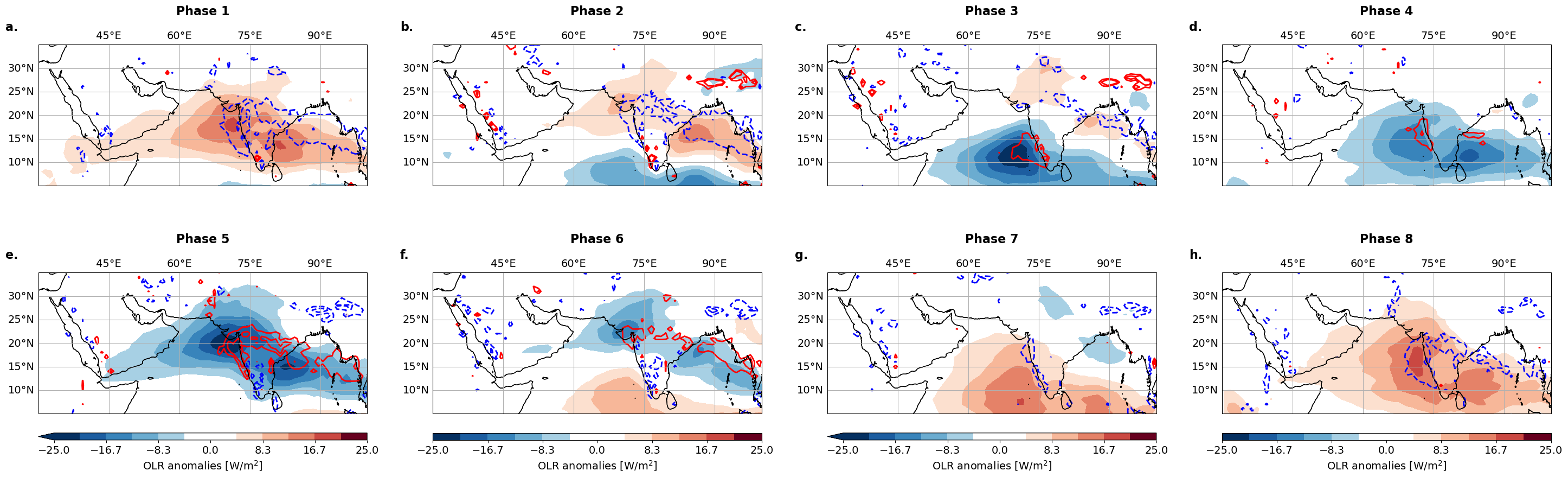}
    \caption{\textbf{Convection centers for BSISO phases.} For phases 1-8 of the BSISO the convection is shown in terms of OLR anomalies (color shading) and vertical velocity $\omega$ in contours where solid red (dashed blue) contours indicate anomalously upward (downward) motion.}
    \label{fig:bsiso_phases_all_tps}
\end{figure}

\section{Quasi-Geostrophic Forcing (QG)} \label{sec:qg_analysis_SI}

\begin{figure}[!htb]
    \centering
    \includegraphics[width=\linewidth]{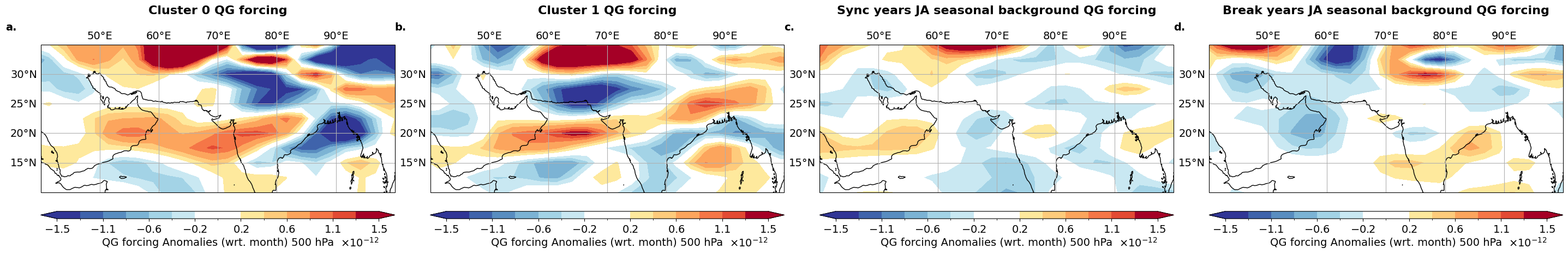}
    \caption{\textbf{Illustration of QG forcing.} Quasi Geostrophic (QG) is visualized for the differences cluster (\textbf{a,b}), for the July and August (JA) mean state for years with many synchronizations (\textbf{c}) and for the July August background state for years with only a few synchronizations, called dry years.
        Only statistically significant anomalies at the 95~\% confidence level are shown.}
    \label{fig:qgforcing}
\end{figure}

The Quasi-Geostrophic (QG) forcing is visualized for the differences cluster, for the July and August (JA) mean state for years with many synchronizations and for the July August background state for years with only a few synchronizations, called dry years (Fig.~\ref{fig:qgforcing}). The QG forcing is calculated using the method described in \cite{Hunt2022} and is based on the anomalies of the geopotential height at $500~$hPa and the wind fields at $200~$hPa.
The QG-forcing is computed to break down the changes in vertical velocity using the (also diagnostic) $\omega$ equation:
\begin{equation}
    \left(\nabla_p^2 + \frac{f}{\sigma}\frac{\partial^2}{\partial p^2}\right) \omega =
    \underbrace{\frac{f_0}{\sigma} \frac{\partial}{\partial p} \left[v_g \nabla_p (\xi_g + f)\right] + \frac{R}{\sigma p} \nabla_p^2\left[v_g \cdot \nabla T\right]}_{\textrm{quasi-geostrophic forcing}} +
    \underbrace{\frac{R}{\sigma p} \nabla_p^2Q}_{\text{local diabatic heating}} \; .
\end{equation}
The symbols have the following meanings: $\omega$ is the vertical velocity, $f$ is the Coriolis parameter, $\sigma$ is the static stability, $v_g$ is the geostrophic wind, $\xi_g$ is the geostrophic vorticity, $R$ is the gas constant, $T$ is the temperature, and $Q$ is the diabatic heating.
Here, the left-hand side can be interpreted as a three-dimensional Laplacian of vertical velocity, acting as a negative operator. On the right-hand side, the first term represents the forcing caused by differential advection of geostrophic absolute vorticity, the second term represents the forcing due to temperature advection, and the third term represents the forcing due to local diabatic heating. The first two terms together form the quasi-geostrophic forcing, which explains the presence of an ascent maximum and precipitation maximum in the southwest quadrant of LPSs. The third term, related to local diabatic heating, is not considered in this context (Hunt, 2022).

\section{Lagrangian Trajectories between North India and Sahel Zone}
\label{sec:lagr_trajs_SI}
\begin{figure}[!htb]
    \centering
    \includegraphics[width=1.\linewidth]{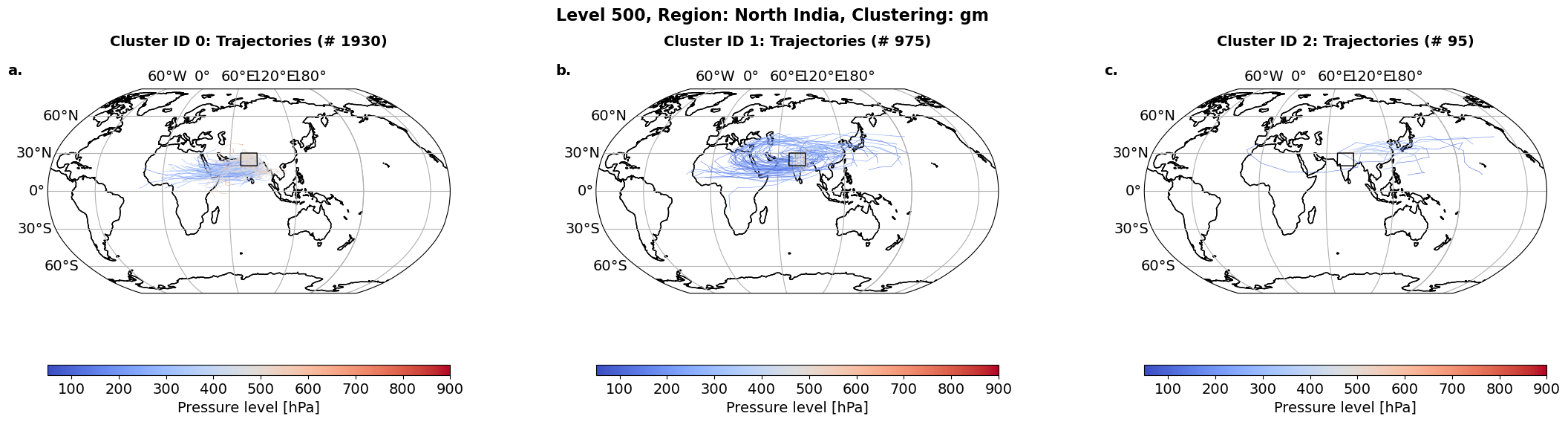}
    \caption{\textbf{Lagrangian Trajectory analysis for trajectories starting in North India.}
        For the most synchronous days in the wet years, we apply a Lagrangian Trajectory analysis in an implementation following the LAGRANTO approach \cite{Sprenger2015} for spatial location in the North India box (here marked by the rectangle). For visual reasons in every group, only every 10th trajectory is plotted.
        The time point for the start of the trajectories is set as the days of maximum synchronization occur and we set the starting pressure level at $500\,$hPa. The resulting trajectories (in total there are around $100$ most synchronous days and 20 starting locations, i.e. $\approx 3000$ trajectories) are clustered using a Gaussian Mixture model algorithm.
        We identify 3 different groups of trajectories.
        \textbf{a} shows the most likely group. Air is rising at NI and then is deflected South- and Westward, merging with the Tropical Easterly Jet (TEJ) around the Sahel Zone.
        \textbf{b} shows similarly the rising air at North India which then propagates towards SZ but is then transported via the subtropical jet stream further to North China.
        We conclude that the propagation of extremes seems to follow the summer mean flow in the upper atmosphere at around $200~$hPa and the TEJ plays a role in the westward propagation.
    }
    \label{fig:lagr_trajs}
\end{figure}

To visualize possible trajectories we use a Lagrangian trajectory model. The Lagrangian trajectory model is based on the LAGRANTO approach \citep{Sprenger2015} and is implemented in the geoutils package \citep{geoutils}.
The trajectories are calculated for the most synchronous days in the wet years. The time point for the start of the trajectories is set as the days of maximum synchronization occur and we set the starting pressure level at $500\,$hPa. The resulting trajectories are clustered using a Gaussian Mixture model algorithm. We identify 3 different groups of trajectories. The most likely group is shown in Fig.~\ref{fig:lagr_trajs}~a. Air is rising at NI and then is deflected South- and Westward, merging with the Tropical Easterly Jet (TEJ) around the Sahel Zone. The second group is shown in Fig.~\ref{fig:lagr_trajs}~b. Air is rising at North India which then propagates towards SZ but is then transported via the subtropical jet stream further to North China. We conclude that the propagation of extremes seems to follow the summer mean flow in the upper atmosphere at around $200~$hPa and the TEJ plays a role in modulating the westward propagation.

\end{document}